\documentclass[12pt]{article}
\usepackage{amsmath}
\usepackage{amssymb} 
\usepackage{bbm} 
\usepackage[small]{caption2} 
\usepackage{fleqn} 
\usepackage{graphicx} 
\usepackage{mathrsfs}   
\usepackage[small,loose]{subfigure}  
\usepackage{cite} 
\usepackage{color}

\newcommand{\Apref}[1]{Appendix~\ref{#1}}
\newcommand{\Secref}[1]{Section~\ref{#1}}
\newcommand{\Eqref}[1]{Equation~\eqref{#1}}
\newcommand{\Figref}[1]{Figure~\ref{#1}}
\newcommand{\Tabref}[1]{Table~\ref{#1}}

\newcommand{\eVdist}{\kern-0.06em}

\newcommand{\ev}{\:\text{e\eVdist V}}   

\newcommand{\gev}{\:\text{Ge\eVdist V}}

\newcommand{\rep}[1]{\ensuremath\boldsymbol{#1}}
\newcommand{\crep}[1]{\ensuremath\overline{\boldsymbol{#1}}}
\DeclareMathOperator{\re}{Re}
\DeclareMathOperator{\im}{Im}
\DeclareMathOperator{\tr}{tr}
\DeclareMathOperator{\Tr}{Tr}

\DeclareMathOperator{\adj}{adj}

\newcommand{\D}{\mathrm{d}}
\newcommand{\I}{\mathrm{i}}

\newcommand{\E}[1]{\ensuremath{\mathrm{E}_{#1}}} 

\newcommand{\SO}[1]{\ensuremath{\mathrm{SO}(#1)}}
\newcommand{\SU}[1]{\ensuremath{\mathrm{SU}(#1)}}
\newcommand{\U}[1]{\ensuremath{\mathrm{U}(#1)}}
\newcommand{\Z}[1]{\ensuremath{\mathbbm{Z}_{#1}}} 

\newcommand{\hu}{\ensuremath{H_u}}
\newcommand{\hd}{\ensuremath{H_d}}
\newcommand{\qhu}{\ensuremath{q_{\hu}}}
\newcommand{\qhd}{\ensuremath{q_{\hd}}}
\newcommand{\singlet}{\ensuremath{N}}
\newcommand{\dilaton}{\ensuremath{S}}
\newcommand{\ie}{i.e.} 
\newcommand{\cf}{cf.}
\newcommand{\Lagrangean}{Lagrange density}

\numberwithin{equation}{section}
\numberwithin{table}{section}

\allowdisplaybreaks[1]

\title{Discrete $\boldsymbol{R}$ symmetries for the MSSM
and its singlet extensions
}

\begin{document}

\begin{titlepage}

\vspace*{-3.0cm}
\begin{flushright}
TUM-HEP 793/11; LMU-ASC 06/11; OHSTPY-HEP-T-11-001; CERN-PH-TH/2011-022;
OUTP-11-33P
\end{flushright}


\begin{center}
{\Large\bf
Discrete $\boldsymbol{R}$ symmetries for the MSSM\\
and its singlet extensions
}

\vspace{1cm}

\textbf{
Hyun Min Lee$^a$,
Stuart Raby$^b$,
Michael Ratz$^c$,\\
Graham G.~Ross$^{a,d}$,
Roland Schieren$^c$,
Kai Schmidt-Hoberg$^d$,\\
Patrick~K. S. Vaudrevange$^e$
}
\\[8mm]
\textit{$^a$\small
~Theory Group, CERN, 1211 Geneva 23, Switzerland
}
\\[5mm]
\textit{$^b$\small
~Department of Physics, The Ohio State University,\\
191 W.\ Woodruff Ave., Columbus, OH 43210, USA
}
\\[5mm]
\textit{$^c$\small
~Physik--Department T30, Technische Universit\"at M\"unchen, \\
James--Franck--Stra\ss e, 85748 Garching, Germany
}
\\[5mm]
\textit{$^d$\small
~Department of Physics, Theoretical Physics, University of Oxford, \\
1 Keble Road, Oxford OX 1 3NP, U.K.
}
\\[5mm]
\textit{$^e$\small
~Arnold Sommerfeld Center for Theoretical Physics,\\
Ludwig--Maximilians--Universit\"at M\"unchen, 80333 M\"unchen, Germany
}
\end{center}

\vspace{1cm}

\begin{abstract}
We determine the anomaly free discrete $R$ symmetries, consistent with the
MSSM,  that commute with SU(5) and suppress the $\mu$ parameter and nucleon
decay. We show that the order $M$ of such $\Z{M}^R$ symmetries has to divide 24
and identify 5 viable symmetries.  The simplest possibility is a $\Z{4}^R$
symmetry which commutes with SO(10).  We present a string--derived model with
this $\Z{4}^R$ symmetry and the exact MSSM spectrum below the GUT scale; in this
model $\Z{4}^R$ originates from the Lorentz symmetry of compactified
dimensions. 
We extend the discussion to include the singlet extensions of the MSSM and find
$\Z4^R$ and $\Z8^R$ are the only possible symmetries capable of solving the
$\mu$ problem in the NMSSM. We also show that a singlet extension of the MSSM
based on a $\Z{24}^R$ symmetry can provide a simultaneous solution to the $\mu$
and strong CP problem with the axion coupling in the favoured window. 
\end{abstract}

\end{titlepage}

\newpage

\section{Introduction}

Supersymmetric extensions of the standard model (SM), such as the minimal
supersymmetric extension, the MSSM, promise to eliminate the hierarchy problem.
However they also introduce serious potential problems and to be viable they
must evade the $\mu$--problem and the problem associated with new baryon- and
lepton--number violating terms. This suggests that there should be an additional
underlying symmetry capable of controlling these terms. Dangerous dimension four
operators can be forbidden by $R$- or matter parity
\cite{Farrar:1978xj,Dimopoulos:1981zb,Dimopoulos:1981dw}, which is an anomaly
free \Z2 subgroup of the continuous baryon minus lepton symmetry $\U1_{B-L}$.
Dimension five proton decay operators can be forbidden by `baryon triality'
\cite{Ibanez:1991pr}, which combines with matter parity to `proton hexality'
\cite{Babu:2003qh,Dreiner:2005rd}. The latter is the unique anomaly free
discrete non--$R$ symmetry forbidding the dangerous operators while allowing the
usual Yukawa couplings, the $\mu$-term and the effective neutrino mass operator.
Anomaly freedom is believed to be a necessary property of discrete symmetries as
otherwise quantum gravity effects may render them inefficient 
\cite{Krauss:1988zc,Ibanez:1991hv,Banks:1991xj,Ibanez:1991pr}.

However, there are two unpleasant properties of these traditional discrete
symmetries. First, they do not allow to address the $\mu$ problem. Second, they
do not commute with the symmetries of the grand unified theories (GUTs) \SU5 or
\SO{10}~\cite{Forste:2010pf}.  In \cite{Babu:2002tx} a discrete $R$ symmetry was
identified which  can address  the $\mu$ problem and commutes with \SO{10}. This
$\Z4^R$ symmetry is anomaly free through cancellation by the Green--Schwarz (GS)
mechanism. In a recent paper \cite{Lee:2010gv} we have shown that this $\Z4^R$ 
is the unique possibility which commutes with \SO{10}, and we have pointed out
that it also solves the problem associated with dimension five proton decay
operators. Furthermore it contains matter parity as a \Z2 subgroup  that is left
unbroken after supersymmetry breaking. 

In this paper we extend the discussion to consider the possible discrete
symmetries of the MSSM which commute with \SU5. As we shall see, there are only
five possibilities with the simplest one being the $\Z4^R$. Our analysis applies
to singlet extensions of the MSSM as well. 

The paper is organized as follows. In \Secref{sec:Classification} we prove that
there are only five (generation independent) discrete $\Z{M}^R$ symmetries which
(i)~commute with \SU5, (ii)~allow the usual Yukawa couplings and dimension five
neutrino mass operator and (iii)~address the $\mu$ and proton decay problems of
the MSSM. \Secref{sec:SimpleZ4R} is dedicated to a more detailed discussion of
the simplest such symmetry,  $\Z4^R$.  We present a globally consistent string
compactification with the exact MSSM spectrum below the compactification scale.
The model exhibits the $\Z4^R$ symmetry, which originates from the Lorentz group
of compactified dimensions. In \Secref{sec:SingletExtensions} we discuss
discrete $R$ symmetries in singlet extensions of the MSSM.  In a theory with the
usual NMSSM couplings the discrete $R$ symmetries can, apart from suppressing
the proton decay rate, provide us with a solution to the NMSSM hierarchy
problem. In a different singlet extension, in which the singlet couples
quadratically to the Higgs bilinear, we will identify a unique discrete $R$
symmetry capable of solving the $\mu$ and strong CP problems simultaneously.
Finally, \Secref{sec:Conclusions} contains our conclusions. In two appendices we
present a re--derivation of discrete anomalies in the path integral approach and
collect anomaly coefficients for discrete $R$ and non--$R$ symmetries.

\section{Discrete symmetries of the MSSM}
\label{sec:Classification}

In this section we discuss discrete symmetries of the MSSM which commute with
\SU5 and can solve the $\mu$ problem. As we shall see, the assumption  that
matter $\Z{M}$ charges commute with \SU5 allows us to restrict possible \Z{M}
symmetries of the MSSM, as well as singlet extensions, to only few
possibilities. We start in \Secref{sec:IssuesNonR} by showing that one cannot
address the $\mu$ problem with  non--$R$ symmetries. In
\Secref{sec:SU5constraints} we then turn to the discussion of discrete $R$
symmetries, for which we prove that the order $M$ has to divide 24. Finally, in
\Secref{sec:ClassificationZN-SU5}, we classify all possible charge assignments.

\subsection{Non--$\boldsymbol{R}$ discrete symmetries}
\label{sec:IssuesNonR}

We start by discussing non--$R$ symmetries. We show that such discrete symmetries that are
consistent with \SO{10} or \SU{5} relations for matter, \ie\ universal charges
for quarks and leptons, cannot forbid the $\mu$ term (\cf\ the similar
discussion in \cite{Hall:2002up}).

Consider a $\Z{M}$ symmetry under which the three generations of $Q$, $\overline{U}$ and
$\overline{E}$ carry discrete charge $q_{\boldsymbol{10}}^g$ while $L$ and $\overline{D}$ carry
$q_{\boldsymbol{\overline{5}}}^g$, where $g$ labels the generation index.
Our  conventions are given in \Apref{app:AnomalyCoefficients}. If the \Z{M}
charges obey  the even stronger \SO{10} relations (\ie\
$q_{\boldsymbol{10}}^g =  q_{\boldsymbol{\overline{5}}}^g$), the following
discussion applies as well. The anomaly coefficients
$A_3:=A_{\SU3_C-\SU3_C-\Z{M}}$, 
$A_2:=A_{\SU2_\mathrm{L}-\SU2_\mathrm{L}-\Z{M}}$, $A_1:=A_{\U1_Y-\U1_Y-\Z{M}}$
and $A_0:=A_{\mathrm{grav}-\mathrm{grav}-\Z{M}}$ are  (\cf\
\Eqref{eq:GeneralAnomalyCoefficients} in \Apref{app:AnomalyCoefficients})
\begin{subequations}\label{eq:A-GUTnon-R-generation-dependent}
\begin{eqnarray}
\label{eq:A-GUTnon-R1-generation-dependent}
 A_3
 & = &
 \frac{1}{2}\sum_{g=1}^3(3\cdot q_{\boldsymbol{10}}^g 
+q_{\boldsymbol{\overline{5}}}^g)
 \;,\\
\label{eq:A-GUTnon-R2-generation-dependent}
 A_2
 & = &
 \frac{1}{2}\sum_{g=1}^3(3\cdot q_{\boldsymbol{10}}^g+q_{\boldsymbol{\overline{5}}}^g)
 +\frac{1}{2}\left(\qhu+\qhd\right)
 \;, \\
 A_1 & = & 
 \frac{1}{2}
 \sum_{g=1}^3\left(
 3\cdot q_{\boldsymbol{10}}^g+q_{\overline{\boldsymbol{5}}}^g\right)
 +\frac{3}{5}\cdot\frac{1}{2}\cdot\left(\qhu+\qhd\right)
 \;,\\
 A_0
 & = & 
 \sum_{g=1}^3 \left(10 \cdot q_{\boldsymbol{10}}^g + 5 \cdot
 q_{\boldsymbol{\overline{5}}}^g
 \right) + 2\qhu + 2\qhd\;,
\end{eqnarray}
\end{subequations}
where the sum runs over the generation indices $g$ and $\qhu$ and $\qhd$ denote
the \Z{M} charges of the up--type and down--type Higgs doublets, respectively.
Anomaly freedom requires 
\begin{equation}\label{eq:UnivR}
 (A_{1\le i\le3}\mod\eta)~=~\frac{1}{24}(A_0\mod\eta)~=~\rho
\end{equation}
with $\rho\ne0$ in the case of GS anomaly cancellation (\cf\ \Eqref{eq:DeltaGS}
in \Apref{app:DiscreteGS}).
Here we define 
\begin{equation}\label{eq:eta}
 \eta~:=~
  \left\{\begin{array}{ll}
    M & \text{for $M$ odd}\;,\\
    M/2 & \text{for $M$ even}\;.
 \end{array}\right. 
\end{equation}
Condition \eqref{eq:UnivR} implies
\begin{equation}
A_2-A_3~=~0 \mod \eta
\end{equation}
and hence, also in the case of generation--dependent $\Z{M}$ charges,
\begin{equation}
\label{eq:muallowed}
\frac{1}{2}\left(\qhu+\qhd\right) ~ =~  0 
 \mod \eta\;.
\end{equation}
On the other hand, the condition that the $\mu$ term is allowed is
\begin{equation}
 \qhu+\qhd~=~0\mod M\;.
\end{equation}
We therefore see that, if we demand \SU5 relations for matter charges, a
non--$R$ \Z{M} symmetry cannot be used to address the $\mu$ problem, even if we
allow for GS cancellation of anomalies.

\subsection{Discrete $\boldsymbol{R}$--symmetries}
\label{sec:SU5constraints}

Having seen that non--$R$ symmetries cannot be used to address the $\mu$
problem, we turn to discuss discrete $R$ symmetries. In this subsection, we
derive constraints on the order $M$ of $\Z{M}^R$ symmetries that can solve the
$\mu$ problem and accommodate the structure of the MSSM.

\subsubsection{A constraint on the order $\boldsymbol{M}$}

After adding the contribution of the gauginos and gravitino  the anomaly
coefficients are
\begin{subequations}
\begin{eqnarray}
 A_3^R 
 & = & 
 \frac{1}{2}\sum_{g=1}^3
 \left(3q_{\boldsymbol{10}}^g+q_{\overline{\boldsymbol{5}}}^g\right)
 -3
 \;,\\
 A_2^R 
 & = & 
 \frac{1}{2}\sum_{g=1}^3
 \left(3q_{\boldsymbol{10}}^g+q_{\overline{\boldsymbol{5}}}^g\right)
 +\frac{1}{2}\left(\qhu+\qhd\right)-5
 \;,\\
 A_1^R 
 & = & 
 \frac{1}{2}\sum_{g=1}^3
 \left(3q_{\boldsymbol{10}}^g+q_{\overline{\boldsymbol{5}}}^g\right)
 +\frac{3}{5}\left[\frac{1}{2}\left(\qhu+\qhd\right)-11\right]
 \;,\\
 A_0^R 
 & = & 
 -21+8+3+1
 +\sum_{g=1}^3\left[10 \,(q_{\boldsymbol{10}}^g-1)
 +5\,(q_{\boldsymbol{\overline{5}}}^g-1)\right]
 \nonumber\\
 & & {}
 +2\, (\qhu+\qhd-2)
 \;,
\end{eqnarray}
\end{subequations}
where $q_{\boldsymbol{10}}$, $q_{\overline{\boldsymbol{5}}}$, $\qhu$ and $\qhd$
denote the $R$ charges of the matter and Higgs superfields, \ie\ matter
fermions and Higgsinos have charges $q-1$. 

In the case $\rho\ne0$, the GS mechanism requires the presence of an axion, such
that  $A_0^R$ is to be amended by the axino/dilatino contribution
($q_{\widetilde{a}}=-1$). 

Subtracting the coefficients from each other leads to the universality
conditions
\begin{subequations}
\begin{eqnarray}
 A_2^R-A_3^R & = & 0\mod\eta 
 \quad  \curvearrowright \quad
 \qhu+\qhd~=~4\mod 2\eta
 \;,\label{eq:anomuniv1a}\\
 A_1^R-A_3^R& = &0\mod\eta 
 \quad\nonumber\\
 & \curvearrowright & \quad
 \frac{3}{5}\,\left[\frac{1}{2}\left(\qhu+\qhd\right)-6\right]~=~0\mod \eta\;.
 \label{eq:anomuniv1b}
\end{eqnarray}
\end{subequations}
\Eqref{eq:anomuniv1a} is equivalent to
\begin{equation}
\frac{1}{2}\left(\qhu+\qhd\right)~=~2+\eta\,\ell
\end{equation}
with an integer $\ell$.
Inserting this into \Eqref{eq:anomuniv1b} yields
\begin{equation}
 \frac{3}{5} \left[\ell\,\eta - 4\right] ~=~ k\, \eta
\end{equation}
with another integer $k$. Altogether we find
\begin{equation}
 \left[3\,\ell-5\,k\right]~=~12/\eta~=~
 \left\{\begin{array}{ll}
  24/M\;,\quad & \text{for $M$ even}\;,\\
  12/M\;,\quad & \text{for $M$ odd}\;.
 \end{array}\right.
\end{equation}
In both cases $24/M $ has to be integer, \ie\ $M$ has to divide 24. Thus the
possible values of $M$ are $3,4,6,8,12$ and $24$.\footnote{We exclude the case
$M=2$ since there are no meaningful order 2 discrete $R$ symmetries (\cf\ e.g.\
\cite{Dine:2009swa}).}  In what follows, we consider all these possibilities.

\subsubsection{Classification}
\label{sec:ClassificationZN-SU5}

Given the constraints on the order $M$, it is straightforward to classify all
phenomenologically attractive charge assignments. Here we assume that the charge assignments are family blind. Though not absolutely necessary it does ensure that the symmetry does not prevent mixing between families in the fermion mass matrix. 
The classification was done by a scan over all possible values of $M$.
In addition to forbidding the $\mu$ term we require that
\begin{enumerate}
 \item Mixed gauge--$\Z{M}^R$ anomalies cancel,
 \ie\ $A_{1\le i\le3}^R=\rho\mod \eta$;
 \item Yukawa couplings $\boldsymbol{10}\,\boldsymbol{10}\,\hu$ and 
 $\boldsymbol{10}\,\overline{\boldsymbol{5}}\,\hd$ as well as the 
 neutrino mass Weinberg operator
 $\overline{\boldsymbol{5}}\,\hu\,\overline{\boldsymbol{5}}\,\hu$ are allowed;
 \item $R$--parity violating couplings are forbidden.
\end{enumerate}
Under these constraints the allowed charge assignments are given in \Tabref{tab:ChargeAssignments}. 
\begin{table}[!h!]
\centerline{
\begin{tabular}{|c|c|c|c|c|c|c|c|c|}
\hline
 $M$ & $q_{\boldsymbol{10}}$ & $q_{\overline{\boldsymbol{5}}}$ & 
 	$\qhu$ & $\qhd$ & $\qhu^\mathrm{sh}$ & $\qhd^\mathrm{sh}$ 
	& $\rho$ & $A_0^{R}(\mathrm{MSSM})$ \\
\hline
 4 & 1 & 1 & 0 & 0 & 16 & 16 & 1 & 1\\
 6 & 5 & 3 & 4 & 0 & 28 & 24 & 0 & 1\\
 8 & 1 & 5 & 0 & 4 & 24 & 28 & 1 & 3\\
 12 & 5 & 9 & 4 & 0 & 28 & 24 & 3 & 1\\
 24 & 5 & 9 & 16 & 12 & 88 & 84 & 9 & 7\\
\hline
\end{tabular}
}
\caption{Phenomenologically attractive charge assignments.  The charges
$\qhu^\mathrm{sh}$ and $\qhd^\mathrm{sh}$  are Higgs charges shifted in such a
way that the anomaly coefficients $A_i^R$ ($1\le i\le 3$) are manifestly
universal. $\rho$ is the universal value of the anomaly coefficients; $\rho\ne0$
indicates GS cancellation of anomalies.}
\label{tab:ChargeAssignments}
\end{table}

For completeness we note that there are only two more charge assignments that are allowed demanding just the first two conditions. 
They are given in \Tabref{tab:AdditionalChargeAssignments}.

\begin{table}[h]
\centerline{%
\begin{tabular}{|c|c|c|c|c|c|c|}
\multicolumn{7}{c}{} \\[-0.2cm]
\hline
$M$ & $q_{\boldsymbol{10}}$ & $q_{\overline{\boldsymbol{5}}}$ & 
 	$\qhu$ & $\qhd$ & $\qhu^\mathrm{sh}$ & $\qhd^\mathrm{sh}$ 
	\\
\hline
 3 & 2 & 0 & 1 & 0 & 10 & 12 \\
 6 & 2 & 0 & 4 & 0 & 28 & 24 \\
\hline
\end{tabular}
}
\caption{Charge assignments which satisfy only the first two criteria. 
Both assignments have $\rho=0$.}
\label{tab:AdditionalChargeAssignments}
\end{table}

One may ask whether there are additional discrete symmetries, such as
$\Z{O}^{(R)}\times\Z{P}^{(R)}$, which cannot be written as single $\Z{M}^{(R)}$
symmetries but also fulfill the three criteria above. The only candidates for
such symmetries are based on the two patterns shown in
\Tabref{tab:AdditionalChargeAssignments}. We find that by amending these
assignments by the usual matter parity one arrives at the $\Z6^R$ symmetry of
\Tabref{tab:ChargeAssignments}. Hence our classification also comprises the
$\Z{O}^{(R)}\times\Z{P}^{(R)}$ case. Of course, in extensions of the MSSM, extra
states can enjoy additional symmetries.

\subsubsection{Dimension five nucleon decay operators}

Note that the third condition is sufficient to eliminate baryon and lepton
number violation due to dimension four terms in the \Lagrangean.  However in the
MSSM at dimension five there are problematic operators allowed that generate
nucleon decay. To be consistent with the bounds on nucleon decay these must be
suppressed by a mass scale more than eight orders above the Planck scale, a
major problem. However in the case of the $\Z{M}^R$ symmetries these operators
are automatically absent. To see this note that the requirement that up- and
down--type Yukawa couplings be allowed implies
\begin{equation}
 3q_{\boldsymbol{10}}+q_{\boldsymbol{\overline{5}}}+\qhu+\qhd~=~4\mod M\;.
\end{equation}
Combining this with \Eqref{eq:anomuniv1a} gives
\begin{equation}\label{eq:dim5automatic}
 3q_{\boldsymbol{10}}+q_{\boldsymbol{\overline{5}}}~=~0 \mod M\;,
\end{equation}
showing that (for $M \neq 2$) the troublesome dimension five operators 
$\boldsymbol{10}\,\boldsymbol{10}\,\boldsymbol{10}\,\overline{\boldsymbol{5}}$
are automatically forbidden whenever the Yukawa couplings are allowed. 

\subsubsection{The gravitational anomaly constraint}

For all charge assignments, the MSSM contribution to the gravitational anomaly is
\begin{equation}
 A_0^{R}(\mathrm{MSSM})~=~7\mod\eta\;.
\end{equation}
All cases except for $M=6$ have $\rho\ne0$ and hence require the presence of an
axion $a$. Call the multiplet containing the axion $\dilaton$,
\begin{equation}
\label{eq:dilatonaxion}
 \dilaton|_{\theta=0}~=~s+\I\,a\;;
\end{equation}
later we will identify $\dilaton$ with the dilaton. From the coupling to the
gauge fields $\int\!\D^2\theta\,S\,W_\alpha\,W^\alpha$ one infers that the
axino/dilatino has $R$ charge $-1$. Therefore, after adding the axino/dilatino
contribution we obtain 
\begin{equation}\label{eq:A0tot}
 A_0^{R}(\mathrm{MSSM}+\mathrm{axino/dilatino})~=~6\mod\eta\;.
\end{equation}
The condition for anomaly freedom is
\begin{equation}\label{eq:GSAnomalyCancellationMSSM}
 \frac{1}{24}\left(A_0^R\mod\eta\right)~=~A_i^R\mod \eta
\end{equation}
for $1\le i\le3$. Now, since $A_i^R \in\Z{}$ and since the order $M$, and
therefore $\eta$, divides 24, this condition is equivalent to 
\begin{equation}
\label{eq:A0R}
 A_0^R~=~0\mod\eta\;.
\end{equation}
From \Eqref{eq:A0tot} we see that the cases $M=4$ and 12 are anomaly free. The
case $M=6$ is anomaly free with an axion that is singlet under $\Z6^R$. All the
other cases require  additional states in order to cancel the gravitational
anomaly.

However this does not necessarily require additional states in the low energy
spectrum. This is because states contributing to the anomaly can acquire mass
when the symmetry is spontaneously broken.  Since the $R$ symmetry is broken in
the hidden sector when supersymmetry is broken these states can acquire a mass
of  order the supersymmetry breaking scale in the hidden sector which can be as
large as $10^{13} \gev$. With this in mind we will not consider the
gravitational anomaly any further.

\subsubsection[Compatibility with $\SO{10}$]{Compatibility with $\boldsymbol{\SO{10}}$}

By looking at the symmetries in \Tabref{tab:ChargeAssignments} we observe that
only the $\Z4^R$ symmetry is compatible with a complete unification of  quarks
and leptons ($q_{\rep{10}}=q_{\crep{5}}$). We will now show that also the other
cases can potentially be in accordance with \SO{10}.  The $\Z{M}^R$ can be a
mixture of the $\U1_X$ subgroup of \SO{10} and an additional
$\widetilde{\mathbbm{Z}}_M^R$ symmetry which commutes with \SO{10}.

The crucial point is to realize that \SO{10} has rank five, and \SU{5} and the
SM gauge group have rank four. Hence, there is an extra $\U1_X$ factor.  We will
denote the $\U1_X$ charge by $Q_X$. The branching rules are
\cite{Slansky:1981yr}
\begin{subequations}
\begin{eqnarray}
\SO{10}  & \supset & \SU{5} \times \U1_X\;, \\
\rep{10} & \rightarrow & \rep{5}_{2} + \crep{5}_{-2}\;, \\
\rep{16} & \rightarrow & \rep{10}_{-1} + \crep{5}_3 +\rep{1}_{-5} \;.
\end{eqnarray}
\end{subequations}
Consider an \SO{10} GUT with an additional $\widetilde{\mathbbm{Z}}_M^R$
symmetry as given in \Tabref{tab:Z_N_From_SO10_general}.
\begin{table}[!h!]
\begin{center}
\subtable[MSSM field content.]{
\begin{tabular}{ccc}
Label & \SO{10} & $\widetilde{\mathbbm{Z}}_M^R$\\
$M_i$ & $\rep{16}$ & $r_{M_i}$\\
$H$ & $\rep{10}$ & $r_H$\\
\end{tabular}
}
\qquad
\subtable[Higgs sector.]{
\begin{tabular}{ccc}
Label & \SO{10} & $\widetilde{\mathbbm{Z}}_M^R$\\
$\psi_H$ & $\rep{16}$ & $r_{\psi_H}$ \\
$\psi_{\overline{H}}$ & $\crep{16}$ & $r_{\psi_{\overline{H}}}$ \\
\end{tabular}
}
\\[5mm]
\subtable[$\widetilde{\mathbbm{Z}}_M^R$--charges for different values of $M$,
the resulting $\Z{M}^R$--charge as a linear combination of the $\U1_X$ charge
$Q_X$  and the $\widetilde{\mathbbm{Z}}_M^R$ charge $r$. In the last column we
list the value of the \SO{10}--\SO{10}--$\widetilde{\mathbbm{Z}}_M^R$ anomaly
$\rho$.]{\begin{tabular}{ccccccc}
$M$ & $r_{M_i}$ & $r_H$ & $r_{\psi_H}$ & $r_{\psi_{\overline{H}}}$ & $\Z{M}^R$ & $\rho$\\
6 & 3 & 2 & 2 & 4 & $5\left(-4Q_X+5r\right)$ & 2\\
8 & 4 & 2 & 7 & 1 & $3\left(Q_X-5r\right)$ & 3\\
12 & 9 & 8 & 8 & 4 & $5\left(-4Q_X+5r\right)$ & 5\\
24 & 12 & 2 & 23 & 1 & $7\left(Q_X-5r\right)$ & 11\\ 
\end{tabular}
}
\end{center}
\caption{Field content of an $\SO{10}\times\widetilde{\mathbbm{Z}}_M^R$ model which can produce the 
symmetries in \Tabref{tab:ChargeAssignments}.}
\label{tab:Z_N_From_SO10_general}
\end{table}
If the \SU{5} singlets contained in the $\rep{16}$ and $\crep{16}$
representations, $\psi_H$ and $\psi_{\overline{H}}$, attain VEVs, which have
$\U1_X$ charge $Q_X=\pm 5$, we arrive at the following breaking pattern
\begin{equation}
 \SO{10} \times \widetilde{\mathbbm{Z}}_M^R \rightarrow \SU{5} \times \Z{M}^R\:.
\end{equation}
Note that there appears to be an additional \Z{5} symmetry, which does however
not constrain any couplings since it is the non--trivial center of \SU5 (\cf\
\cite{Csaki:1997aw,Petersen:2009ip}).

In summary, we can obtain our $\Z{M}^R$ symmetries from an \SO{10} GUT. However,
the scenarios presented here are only toy models. First of all,  further Higgs
fields are needed to break \SU5 down to the SM. In addition, to obtain
doublet--triplet splitting and get rid of dimension five operators larger  Higgs
representations are needed. Also, anomaly matching (\cf\ \cite{Csaki:1997aw})
forces us to introduce extra representations because the value of $\rho$  does
not equal the one given in \Tabref{tab:ChargeAssignments}.

These considerations also show that the \SU{5} relations are mandatory. Since
\SU{5} and the standard model gauge group have the same rank,  there is no \U1
with which our $\Z{M}^R$s could mix upon breaking \SU5.


\section{A simple $\boldsymbol{\Z4^R}$ symmetry in the MSSM}
\label{sec:SimpleZ4R}

In \Tabref{tab:ChargeAssignments} we survey all symmetries and charge
assignments which commute with \SU5. The simplest one, the $\Z4^R$, commutes
also with \SO{10}. In what follows we will discuss this case in more detail.

\subsection{Non--perturbative terms}

The gauge invariant superpotential of the MSSM contains
\begin{eqnarray}
\lefteqn{\mathscr{W} ~=~ 
\mu\, \hu\,\hd + \kappa_i\, L_i \hu} \nonumber \\
& &{}+ Y_e^{ij}\hd\, L_i\,  \overline{E}_j + Y_d^{ij}\hd\, Q_i\,\overline{D}_j 
+ Y_u^{ij}\hu\, Q_i\,\overline{U}_j 
\nonumber\\
& &{} + \lambda^{(0)}_{ijk}\, L_i L_j \overline{E}_k 
+ \lambda^{(1)}_{ijk}\, L_i Q_j \overline{D}_k  
+ \lambda^{(2)}_{ijk}\, \overline{U}_i\,\overline{D}_j\,\overline{D}_k 
\nonumber\\
& &{} 
+\kappa^{(0)}_{ij}\, \hu\, L_i\,\hu\, L_j
+\kappa^{(1)}_{ijk\ell}\, Q_i\, Q_j\, Q_k\, L_\ell 
+ \kappa^{(2)}_{ijk\ell}\,\overline{U}_i\,\overline{U}_j\,\overline{D}_k\, \overline{E}_\ell
\nonumber\\
& &{} 
+\kappa^{(3)}_{ijk}\,Q_i\, Q_j\, Q_k\, \hd
+\kappa^{(4)}_{ijk}\,Q_i\,\overline{U}_j \, \overline{E}_k \, \hd
+\kappa^{(5)}_{i}L_i \hu \hu \hd
\;.\label{eq:GeneralWMSSM}
\end{eqnarray}
We see immediately that the coefficients $\mu$, $\kappa_i$,
$\lambda^{(0)}_{ijk}$, $\lambda^{(1)}_{ijk}$, $\lambda^{(2)}_{ijk}$,
$\kappa^{(1)}_{ijk\ell}$, $\kappa^{(2)}_{ijk\ell}$, $\kappa^{(3)}_{ijk}$,
$\kappa^{(4)}_{ijk}$ and $\kappa^{(5)}_i$  are forbidden by $\Z4^R$
perturbatively while $Y_{e,d,u}^{ij}$ and $\kappa^{(0)}_{ij}$ are allowed.  In
what follows we will show that at the non--perturbative level $\mu$ as well as
$\kappa^{(1)}_{ijk\ell}$ and $\kappa^{(2)}_{ijk\ell}$ will be induced while the
$R$ parity violating couplings $\kappa_i$ and $\lambda$ as well as the
$\kappa^{(3-5)}$ remain zero. The reason is that the latter are forbidden by a
\Z2 subgroup of $\Z4^R$ which is equivalent to matter parity.  This subgroup is
unbroken by the supersymmetry breaking sector and thus remains a symmetry of the
full theory.

Let us spell out the argument in somewhat more detail. Call the $\Z4^R$
transformation $\zeta$,
\begin{eqnarray}
 \zeta~:~
 \text{matter superfield}& \to & \I\cdot\text{matter superfield}\;,\nonumber\\
 \text{Higgs superfield}& \to & \text{Higgs superfield}\;,\nonumber\\
 \theta& \to &\I\cdot\theta\;,\nonumber\\
 \mathscr{W} & \to & -\mathscr{W}\;.
\end{eqnarray} 
Now look at the transformation $\zeta^2$, under which matter superfields
transform with a minus,  Higgs superfields go into themselves and $\theta
\to-\theta$. The transformation $\text{fermion}\to-\text{fermion}$ and $\theta
\to - \theta$ is a symmetry of any SUSY theory, therefore $\zeta^2$ is
equivalent to matter parity, and, in particular, anomaly free with $\rho=0$. 
One can use the path integral (\cf\ \Apref{app:PathIntegral}) to show that
correlators that vanish due to a non--anomalous symmetry with $\rho=0$ also
vanish at the quantum level. Therefore, the matter parity subgroup contained in
the $\Z4^R$ will not be violated by quantum effects.

On the other hand, correlators which are only forbidden by $\Z4^R$ but
not by $\Z2$, \ie\ which are invariant under $\zeta^2$, can be
non--trivial at the quantum level. A convenient way to parametrize effective
couplings describing these effects involve the \dilaton\ field, which shifts
under the $\Z4^R$ symmetry as (\cf\ \Eqref{eq:DiscreteShiftDilaton})
\begin{equation}
 \dilaton~\to~\dilaton+\frac{\I}{2}\Delta_\mathrm{GS}\;.
\end{equation}
The discrete shift of $\dilaton$ is given by (\cf\ \Eqref{eq:DeltaGS})
\begin{equation}
 \Delta_\mathrm{GS}~=~\frac{1}{4\pi}(A_{G-G-\Z{4}^R}\mod2)
 ~=~\frac{1+2\nu}{4\pi}
\end{equation}
with $\nu\in\mathbbm{Z}$.
This allows us to write down terms 
\begin{eqnarray}
 \Delta\mathscr{W}_\mathrm{np}
 & = &
 \exp\left(-8\pi^2\,\frac{1+2n}{1+2\nu}\,\dilaton\right)\,\left[
 B_0+ \overline{\mu}\,\hu\,\hd
 +\overline{\kappa}^{(1)}_{ijk\ell}\, Q_i\, Q_j\, Q_k\, L_\ell 
 \right.
 \nonumber\\
 & & \left.
 \phantom{\exp\left(-8\pi^2\,\frac{}{1+2\nu}\,\dilaton\right)\,\left[\right.}{}
 +\overline{\kappa}^{(2)}_{ijk\ell}\,\overline{U}_i\,\overline{U}_j\,\overline{D}_k\, \overline{E}_\ell
 \right]\label{eq:DeltaWnp}
\end{eqnarray}
with some coefficients $B_0$, $\overline{\mu}$ and
$\overline{\kappa}^{(1,2)}_{ijk\ell}$ and $n\in\mathbbm{Z}$. Such superpotential
terms are $\Z4^R$ covariant, \ie\ the exponential transforms with a minus under
$\Z4^R$ while the terms in the square brackets are invariant. Due to the fact
that $\dilaton$ enters the gauge kinetic function, these terms are proportional
to $\mathrm{e}^{-8\pi^2\frac{1+2n}{1+2\nu}\,\frac{1}{g^2}}$. For $n=\nu=0$ they
can be interpreted as originating from t'Hooft instanton effects. The $8\pi^2$
in the exponential can also be obtained directly in a stringy
computation~\cite{Kappl:2010yu}. The crucial property of the non--perturbative
couplings \eqref{eq:DeltaWnp} is that they are naturally suppressed.

The critical question concerns now the interpretation of the
$\mathrm{e}^{-8\pi^2\frac{1+2n}{1+2\nu}\,S}$ terms. So far we have shown that
such terms are $\Z4^R$ covariant. In the MSSM as a `stand--alone' theory,
$\SU3_C$ or $\SU2_\mathrm{L}$ instantons can generate such terms, but their
magnitude turns out to be very small. Whether or not further terms, with given
$n$ and $\nu$, appear depends on the model. Let us now make the very common
assumption that there is a hidden sector that gets strong at some intermediate
scale $\Lambda$.  Then the non--perturbative terms related to the strong
dynamics may well be the source of supersymmetry breakdown
\cite{Witten:1981nf,Nilles:1982ik}. Given non--renormalizable interactions
between the MSSM and the hidden sector, communicated by some messenger fields, 
$\Lambda$ sets the magnitude of the MSSM soft terms,
$m_\mathrm{soft}\sim\Lambda^3/M_*^2$, with $M_*$ being the messenger scale. In
such settings,  holomorphic, \ie\ superpotential, terms can also be induced by
higher--dimensional operators. That is, the $\Delta\mathscr{W}_\mathrm{np}$
terms can appear with magnitude $m_\mathrm{soft}\sim\Lambda^3/M_*^2$, but in
principle they may also be absent if there are no higher--dimensional operators
connecting the MSSM sector with the hidden sector exhibiting strong dynamics. In
other words, if the MSSM fields are singlets under the hidden sector gauge
interactions, there is, a priori, no guarantee that the
$\Delta\mathscr{W}_\mathrm{np}$ terms appear with reasonable size.  If the scale
of MSSM soft terms is related to some hidden sector strong dynamics, we expect
the holomorphic terms also to appear, unless there are additional symmetries
beyond $\Z4^R$ that forbid such couplings. Assuming that the dominant
non--perturbative scale is related to supersymmetry breakdown we expect that the
$\Delta\mathscr{W}_\mathrm{np}$ terms are of the order of the soft supersymmetry
breaking terms. We will mainly focus on gravity mediation, where
$M_*=M_\mathrm{P}$ and these terms are of the order of the gravitino mass
$m_{3/2}$ (in Planck units). Below in \Secref{sec:StringExample} we will present
an explicit string theory example in which the non--perturbative $\mu$ term is
directly connected to $m_{3/2}$. 

At this point let us mention that for the case of discrete $R$ symmetries we
disagree with statements made recently in~\cite{Dine:2009swa}, where it was
claimed that, in the context of gravity mediation, $R$ symmetries will be broken
at the Planck scale and be therefore ineffective. The claim relies on the
observation that there are fields with Planck scale VEVs that break the $R$
symmetry.  The derivation of this result relies on the inequality  $|\langle
\mathscr{W}\rangle| \le \tfrac{1}{2} f_r |F|$ (\cf~Equation~(9)
in~\cite{Dine:2009swa}) where $f_{r}$ is the $R$--axion decay constant. This was
derived for the case of continuous $R$ symmetries by taking the limit of an
infinitesimal transformation \cite{Dine:2009swa}. For the case of discrete $R$
symmetries the inequality is no longer true and there is no requirement  that
$R$--non singlets acquire Planck scale VEVs. In this case the $R$ symmetry can
be broken at a much lower scale. This is the case in the supergravity examples
discussed here. In them the breaking of the $R$ symmetry occurs
non--perturbatively  at an intermediate scale in a hidden sector and it is the
superpotential VEV $\langle\mathscr{W}\rangle$ rather than a field VEV that is
the order parameter for $R$ symmetry breaking.  Since the superpotential only
appears at the non--perturbative level it is small. Also all other $R$ symmetry
breaking terms are small.  This applies also to other schemes such as the one
discussed in~\cite{Kappl:2008ie}, where a small $\langle\mathscr{W}\rangle$ is a
consequence of an approximate $R$ symmetry. Here the $R$ symmetry is broken
perturbatively, but again the order parameter, \ie\ the superpotential VEV, is
very small. In conclusion, $R$ symmetries are a useful tool also, or in
particular, in gravity mediation, where the same parameter, the small
superpotential VEV, both sets the scale of soft masses and cancels the vacuum
energy.
In what follows, we discuss how the connection between the
$\Delta\mathscr{W}_\mathrm{np}$ terms and $m_{3/2}$ arises in the scheme of
K\"ahler stabilization.

\subsection{Dilaton stabilization and supersymmetry breaking}

At the present stage of the discussion, the $\dilaton$ field has no potential
and supersymmetry is unbroken. An economical way to rectify this situation is to
invoke the stringy scheme of K\"ahler stabilization
\cite{Shenker:1990uf,Banks:1994sg,Casas:1996zi,Binetruy:1996nx} (see also
\Apref{app:KaehlerStabilization}).\footnote{Alternatively, other stabilization
schemes, such as racetrack mechanisms, may be applicable here.}
In this case the term of the form $\mathrm{e}^{-b\,\dilaton}$ represents a
hidden sector gaugino condensate \cite{Nilles:1982ik}, which sets the scale for
supersymmetry breakdown. According to the above discussion, in the presence of
our $\Z4^R$ symmetry 
\begin{equation}
 b~=~8\pi^2\,\frac{1+2n}{1+2\nu}\;.
\end{equation} 
Let us discuss what that means in the case of a hidden $\SU{N_c}$ theory with
$N_f$ chiral superfields in the $\boldsymbol{N_c}+\overline{\boldsymbol{N_c}}$
representations. Here the coefficient $b$ is given by
\begin{equation}
 b~=~\frac{3}{2\beta}~=~\frac{3\cdot8\pi^2}{3N_c-N_f}\;.
\end{equation}
Therefore
\begin{equation}
 \frac{3}{3N_c-N_f}~=~\frac{1+2n}{1+2\nu}\;.
\end{equation}

In the scheme under consideration, supersymmetry is broken by a non--trivial VEV
of $F_S$. This leads to gaugino and soft scalar masses, following the pattern of
the so--called ``dilaton dominated scenario'' \cite{Kaplunovsky:1993rd}. This
scenario has a number of phenomenologically attractive features. In particular,
due to flavour universality in the soft breaking sector, it avoids the SUSY FCNC
problem. Also, most of the physical CP phases, e.g.\ $\arg(A^* M)$, vanish which
ameliorates the SUSY CP problem.   However  in the dilaton dominated case the
vacuum structure may favour an unacceptable colour  breaking minimum
\cite{Casas:1996wj}. Other phenomenological aspects have been discussed in
\cite{Abel:2000bj}.

Moreover, the (non--perturbative) superpotential acquires a non--trivial VEV as
well,
\begin{equation}
 \langle\mathscr{W}\rangle~\sim~\mathrm{e}^{-b\,\langle\dilaton\rangle}~\ne~0\;.
\end{equation}   
All gauge invariant terms which have been forbidden because they have zero $R$
charge can now be obtained by multiplying them with $\langle\mathscr{W}\rangle$.
$\langle\mathscr{W}\rangle$ will hence be the order parameter for $R$ symmetry
breaking. Inserting this in \Eqref{eq:DeltaWnp} we find that there will be a
$\mu$ term of the order of $\langle\mathscr{W}\rangle$, \ie\ of the order of the
gravitino mass $m_{3/2}$, as well as
$\kappa^{(1)}_{ijk\ell}\sim10^{-15}/M_\mathrm{P}$. On the other hand, terms
which have odd $\Z4^R$ charge cannot be obtained by multiplying them by
$\mathrm{e}^{-b\,\langle\dilaton\rangle}$; these are precisely the $R$ parity
violating couplings $\kappa_i$, $\lambda^{(0)}$, $\lambda^{(1)}$ and
$\lambda^{(2)}$ in \Eqref{eq:GeneralWMSSM}, showing again that matter parity
will not be broken.

\subsection{Phenomenology}

The suppression of the $\kappa^{(1)}$ term leads to a situation in which
dimension five proton decay will be unobservably small. Therefore, proton decay
will proceed through dimension six operators mediated by gauge boson exchange.

In settings with discrete $R$ symmetries one should worry about the cosmological
domain wall problem~\cite{Abel:1995wk}.  The domain walls form at the stage of
$R$ symmetry breaking, typically the scale of supersymmetry breaking. For the
case of gravity mediation this is at an intermediate scale of
$\mathcal{O}(10^{12})\,\mathrm{GeV}$. Provided the Hubble scale during inflation
is below this scale, domain walls have sufficient time to form and then they
will be inflated away. The requirement that no domain walls are created after
inflation translates in an upper bound on the reheat temperature $T_R$, which,
given the other bounds on $T_R$ in supersymmetric cosmology, appears rather
mild. 

A discrete $R$ symmetry may also be useful for inflationary scenarios.  For
example, in \cite{Kumekawa:1994gx}, it is argued that a $\Z8^R$ symmetry,
with inflaton field $\phi$ carrying $R$ charge 2, can be used to guarantee that
the inflaton potential is flat near the origin and give enough
inflation.\footnote{Note, their definition of the order of the discrete
symmetry differs from ours. What they call $\Z4^R$ we call $\Z8^R$.}

In summary,  for the case of gravity mediated supersymmetry breaking,
non--perturbative effects naturally generate a $\mu$ parameter of the order of
the gravitino mass. The symmetry ensures that the proton decay rate is well
below the experimental limit and an exact matter parity is left that guarantees
SUSY particles can only be pair produced and the lightest SUSY particle is
stable. Thus one is left with the usual MSSM phenomenology with negligibly small
corrections from higher dimension terms.

\subsection{$\boldsymbol{\Z4^R}$ literature}

A version of the $\Z4^R$ symmetry has been prosed by Kurosawa et
al.~\cite{Kurosawa:2001iq} where the traditional version of anomaly constraints
was imposed, \ie\ the possibility of GS anomaly cancellation has not been taken
into account. This lead to a setting in which extra light charged states were
required to cancel the anomaly.  

The $\Z4^R$ symmetry with GS anomaly cancellation has also been discussed by
Babu et al.~\cite{Babu:2002tx}.  There are several aspects in which our
analysis differs from or goes beyond \cite{Babu:2002tx}:

\begin{enumerate}
\item We discuss the uniqueness for the first time.

\item We point out that the $\Z4^R$ symmetry also suppresses dimension five
operators.

\item We present the first discussion of non--perturbative $\Z4^R$ violating
effects.

\item Related to the previous point, Babu et al.\ only discuss generation of the
$\mu$ term by the Giudice--Masiero mechanism. However, there will also be a
holomorphic non--perturbative (\ie\ Kim--Nilles type) contribution.

\item In \cite{Babu:2002tx} the $\Z4^R$ is argued to originate from an
`anomalous' $\U1_R$. We are not aware of a model in which such an `anomalous'
$\U1_R$ appears in string models. However, we also cannot rule out this
possibility.

\item We present a detailed discussion of  the mixed hypercharge coefficient
$A_1$.

\item Babu et al.\ do not  discuss the gravitational anomalies.
\end{enumerate}

\subsection{String theory realization}
\label{sec:StringExample}

In the above discussion we argued that, if some hidden sector strong dynamics
was responsible for supersymmetry breakdown, also a $\mu$ term of the right size
will be induced by this dynamics. In order to render our discussion more
specific, we will now discuss an explicit, globally consistent string--derived
model. Such models have the important property that they are complete, \ie\
unlike bottom--up (or `local') models they cannot be `amended' by some extra
states or sectors. This allows us to clarify whether or not a reasonable $\mu$
term will appear.

Making extensive use of the methods to determine the remnant symmetries
described in \cite{Petersen:2009ip}, we were able to find examples realising the
$\Z4^R$ discussed in \Secref{sec:SimpleZ4R}, based on the string model derived
in \cite{Blaszczyk:2009in} and similar models, with the exact MSSM spectrum, a
large top Yukawa coupling, a non--trivial hidden sector etc. In what follows, we
present an explicit example.

\begin{table}[t]
\centerline{
\begin{tabular}{|r|c|l|c|r|c|l|}
\hline
\# & representation & label & & \# & representation & label\\
\hline
$3$ & $(\boldsymbol{3}, \boldsymbol{2}; \boldsymbol{1}, \boldsymbol{1}, \boldsymbol{1})_{ \frac{1}{6}}$            & $Q$ 
& $\phantom{I^{I^I}}$&
$3$ & $\phantom{I^{I^I}}$ $(\boldsymbol{\overline{3}}, \boldsymbol{1}; \boldsymbol{1}, \boldsymbol{1}, \boldsymbol{1})_{-\frac{2}{3}}$ & $\overline{U}$ \\
$8$ & $\phantom{I^{I^I}}$ $(\boldsymbol{\overline{3}}, \boldsymbol{1}; \boldsymbol{1}, \boldsymbol{1}, \boldsymbol{1})_{ \frac{1}{3}}$ & $\overline{D}$
&&
$5$ & $(\boldsymbol{3}, \boldsymbol{1}; \boldsymbol{1}, \boldsymbol{1}, \boldsymbol{1})_{-\frac{1}{3}}$            & $D$ \\
$7$ & $(\boldsymbol{1}, \boldsymbol{2}; \boldsymbol{1}, \boldsymbol{1}, \boldsymbol{1})_{-\frac{1}{2}}$            & $L$
&&
$4$ & $(\boldsymbol{1}, \boldsymbol{2}; \boldsymbol{1}, \boldsymbol{1}, \boldsymbol{1})_{ \frac{1}{2}}$            & $\overline{L}$ \\
$3$ & $(\boldsymbol{1}, \boldsymbol{1}; \boldsymbol{1}, \boldsymbol{1}, \boldsymbol{1})_{ 1}$                      & $\overline{E}$
&&
$33$& $(\boldsymbol{1}, \boldsymbol{1}; \boldsymbol{1}, \boldsymbol{1}, \boldsymbol{1})_{0}$                       & $N$ \\
$5$ & $(\boldsymbol{1}, \boldsymbol{1}; \boldsymbol{3}, \boldsymbol{1}, \boldsymbol{1})_{0}$                       & $X$
&&
$5$ & $(\boldsymbol{1}, \boldsymbol{1};\boldsymbol{\overline{3}}, \boldsymbol{1}, \boldsymbol{1})_{0}$             & $\overline{X}$ \\
$6$ & $(\boldsymbol{1}, \boldsymbol{1}; \boldsymbol{1}, \boldsymbol{1}, \boldsymbol{2})_{0}$                       & $Y$
&&
$6$ & $( \boldsymbol{1}, \boldsymbol{1}; \boldsymbol{1}, \boldsymbol{2}, \boldsymbol{1})_{0}$                      & $Z$ \\
\hline
\end{tabular}
}
\caption{Spectrum of the orbifold model from \cite{Blaszczyk:2009in}. The representations w.r.t. $G_\mathrm{SM}\times[\text{SU}(3)\times\text{SU}(2)\times\text{SU}(2)]_\mathrm{hid}$, their multiplicities (\#) and labels are listed.}
\label{tab:StringZ2xZ2ModelSpectrum}
\end{table}

Consider the MSSM candidate model of \cite{Blaszczyk:2009in}. It is obtained 
by the compactification of the $\E{8}\times\E{8}$ heterotic string on a 
$\Z2\times\Z2$ orbifold with an additional freely acting $\Z2$. At the 
orbifold point (where the VEVs of all fields are set to zero) the 
$\E{8}\times\E{8}$ gauge group gets broken to
\begin{equation}
G_\mathrm{SM}\times[\text{SU}(3)\times\text{SU}(2)\times\text{SU}(2)]_\mathrm{hid}
\end{equation}
times eight $\U1$ factors. One of them, denoted by $\U1_\mathrm{anom}$, appears 
anomalous, \ie\ $\tr Q_\mathrm{anom} = 180 \neq 0$ using the normalization 
$|t_\mathrm{anom}|^2 = 15$. Hence, a one--loop Fayet--Iliopoulos $D$-term gets 
induced. Furthermore, the massless spectrum includes three generations of 
quarks and leptons and is summarized in \Tabref{tab:StringZ2xZ2ModelSpectrum}. 
More details on the model can be found in \cite{Z4RinternetNote:2011da}.

Next, we choose a vacuum configuration in which the SM singlets
\begin{eqnarray}\label{eq:phiFields}
\{\phi_i\} &=&\{\overline{X}_{4},\overline{X}_{5},X_{3},X_{4},X_{5},Y_{1},Y_{2},Z_{1},Z_{2}, \nonumber\\
           & & {} N_{1},N_{2},N_{4},N_{7},N_{10},N_{15},N_{16},N_{17},             \nonumber\\
           & & {} N_{20},N_{21},N_{25},N_{27},N_{28}, N_{30},N_{32},N_{33}\}
\end{eqnarray}
attain VEVs. These fields are charged with respect to the hidden sector gauge 
group, the $\U1$ factors and several discrete symmetries (\ie\ this orbifold 
compactification provides three $\Z4^R$ symmetries reflecting the discrete 
rotational symmetry of the three \Z2 orbifold planes and six $\Z2$ factors coming 
from the space group selection rule, see Appendix B of \cite{Blaszczyk:2009in}). 
Hence, the $\phi_i$ VEVs of \Eqref{eq:phiFields} break these (gauge and discrete) 
symmetries and it turns out that 
\begin{equation}
G_\mathrm{SM}\times\Z4^R\times\Z2\;
\end{equation}
remains unbroken, where $\Z4^R$ is a mixture of an orbifold $\Z4^R$ and other 
symmetries.\footnote{Meanwhile a very similar string model exhibiting vacua without the 
extra \Z2 has been found \cite{Kappl:2010yu}.} The $\Z4^R\times\Z2$ charges of 
the SM charged fields are listed in \Tabref{tab:chargesString}. From there we see 
that the $\Z4^R$ factor gives a stringy realisation of the $\Z4^R$ symmetry 
described in a bottom--up approach in \Secref{sec:SimpleZ4R}. Furthermore, the 
$\phi_i$ VEVs also provide mass terms for the exotics, which are massless at the 
orbifold point, and allow us to cancel the Fayet--Iliopoulos $D$--term. 

\begin{table}[t]
\centerline{%
 \begin{tabular}{c||c}
 quarks and leptons & Higgs and exotics \\
 \hline\hline
 \begin{tabular}{l|c|c|c|l|c|c}
 $ Q_{1}$      & $1$ & $0$ & $\phantom{I^{I^{I^I}}}$ & $ \overline{U}_{1}$ & $1$ & $0$ \\
 $ Q_{2}$      & $1$ & $1$ & & $ \overline{U}_{2}$ & $1$ & $1$ \\
 $ Q_{3}$      & $1$ & $1$ & & $ \overline{U}_{3}$ & $1$ & $1$ \\
  \cline{1-3} \cline{5-7}
 $ \overline{D}_{3}$ & $1$ & $1$ & $\phantom{I^{I^{I^I}}}$ & $ L_{2}$      & $1$ & $1$ \\
 $ \overline{D}_{7}$ & $1$ & $0$ & & $ L_{6}$      & $1$ & $0$ \\
 $ \overline{D}_{8}$ & $1$ & $0$ & & $ L_{7}$      & $1$ & $0$ \\
  \cline{1-3}
 $ \overline{E}_{1}$ & $1$ & $0$ & $\phantom{I^{I^{I^I}}}$ & \multicolumn{3}{c}{} \\
 $ \overline{E}_{2}$ & $1$ & $1$ & & \multicolumn{3}{c}{} \\
 $ \overline{E}_{3}$ & $1$ & $1$ & & \multicolumn{3}{c}{} \\
 \end{tabular}
 &
 \begin{tabular}{l|c|c|c|l|c|c}
 $ \overline{L}_{1}$ & $0$ & $0$ & $\phantom{I^{I^{I^I}}}$ & $ L_{1}$ & $0$ & $0$ \\
 $ \overline{L}_{2}$ & $0$ & $0$ & & $ L_{3}$ & $2$ & $0$ \\
 $ \overline{L}_{3}$ & $2$ & $0$ & & $ L_{4}$ & $0$ & $0$ \\
 $ \overline{L}_{4}$ & $2$ & $0$ & & $ L_{5}$ & $0$ & $0$ \\
  \cline{1-3} \cline{5-7}
 $ \overline{D}_{1}$ & $0$ & $0$ & $\phantom{I^{I^{I^I}}}$ & $ D_{1}$ & $0$ & $0$ \\
 $ \overline{D}_{2}$ & $2$ & $0$ & & $ D_{2}$ & $2$ & $0$ \\
 $ \overline{D}_{4}$ & $2$ & $0$ & & $ D_{3}$ & $0$ & $0$ \\
 $ \overline{D}_{5}$ & $0$ & $0$ & & $ D_{4}$ & $2$ & $0$ \\
 $ \overline{D}_{6}$ & $0$ & $0$ & & $ D_{5}$ & $2$ & $0$ \\
 \end{tabular}
 \end{tabular}
} \caption{$\Z4^R\times\Z2$ charges of the fields with SM quantum numbers.}
\label{tab:chargesString}
\end{table}

In detail, we find four Fayet-Iliopoulos monomials, \ie\ monomials that 
are gauge invariant except for a total negative $\U1_\mathrm{anom}$ 
charge such that their VEVs can cancel the positive Fayet-Iliopoulos 
term in the $D$--term potential (\cf\ \cite{Buccella:1982nx}). The monomials 
read
\[
\big\{N_{28}^4 Y_{1}Y_{2},~
N_{28}^4 Z_{1}Z_{2},~
N_{33}^4 Y_{1}Y_{2},~
N_{33}^4 Z_{1}Z_{2}\big\}\;,
\]
with $Q_\mathrm{anom}(\text{FI monomial}) = -15$. Further, we find 
monomials with zero or positive $\U1_\mathrm{anom}$ charge involving 
all $\phi_i$ fields from \Eqref{eq:phiFields}. Hence, this represents 
a $D$--flat configuration.

Matter fields are identified as fields with $\Z4^R$ charge 1, see 
\Tabref{tab:chargesString}, and are given by
$Q_{1}$, $Q_{2}$, $Q_{3}$,
$\overline{U}_{1}$, $\overline{U}_{2}$, $\overline{U}_{3}$,
$\overline{E}_{1}$, $\overline{E}_{2}$, $\overline{E}_{3}$,
$\overline{D}_{3}$, $\overline{D}_{7}$, $\overline{D}_{8}$,
$L_{2}$, $L_{6}$ and $L_{7}$. An inspection of the discrete charges of the Higgs
candidates, \ie\ the remaining $L$ and $\overline{L}$ fields, reveals that there 
is one massless Higgs pair at the perturbative level. Unfortunately, the
additional \Z2, which we cannot break, forbids some Yukawa couplings such that
the charged lepton and $d$-type Yukawa couplings $Y_e$ and $Y_d$ have rank 2. 

\begin{figure}[!h!]
\centerline{\includegraphics{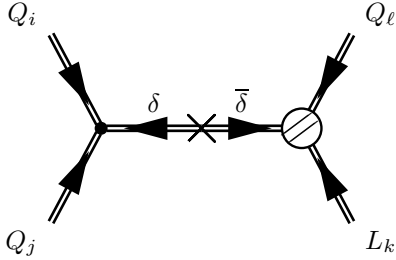}}
\caption{Absence of induced $QQQL$ operators. Either one vertex or the mass
term is forbidden by $\Z4^R$.}
\label{fig:NoDeltaExchange}
\end{figure}

More explicitly, we have computed the couplings using the well-known string selection
rules \cite{Dixon:1986qv,Hamidi:1986vh}, which have been extended to the case of
non-local GUT breaking~\cite{Blaszczyk:2009in}. For the Higgs mass matrix we find
\begin{equation}
 \mathcal{M}_\mathrm{Higgs}~=~
\left(
\begin{array}{cccc}
 0          & N_{15}     & 0       & 0 \\
 0          & \phi ^{13} & 0       & 0 \\
 \phi ^{11} & 0          & N_{10}  & \phi^3 \\
 \phi ^{11} & 0          & \phi^3 & N_{10} 
\end{array}
\right)\;,
\end{equation}
where for example $\phi^3$ denotes a sum of known monomials in the fields of 
\Eqref{eq:phiFields} starting at degree three. Therefore a linear combination
of $\overline{L}_{1}$ and $\overline{L}_{2}$ as well as a linear combination of  $L_{1}$,
$L_{4}$ and $L_{5}$ remains massless. The mass matrix of the extra colour triplets 
\begin{subequations}
\begin{eqnarray}
 \{\delta_i\} & = &\left\{D_{1},D_{2},D_{3},D_{4},D_{5}\right\}\;,
 \\
 \{\overline{\delta}_i\} & = &
 \left\{\overline{D}_{1},\overline{D}_{2},\overline{D}_{4},
 	\overline{D}_{5},\overline{D}_{6}\right\}    
\end{eqnarray}
\end{subequations}
is
\begin{equation}
 \mathcal{M}_\mathrm{extra~triplets}~=~
\left(
\begin{array}{ccccc}
 0      & N_{1}   & \phi ^9    & 0       & 0 \\
 N_{2}  & 0       & 0          & N_{16}  & N_{20}  \\
 0      & \phi ^3 & \phi ^{13} & 0       & 0 \\
 N_{28} & 0       & 0          & N_{10}  & \phi ^3 \\
 N_{33} & 0       & 0          & \phi ^3 & N_{10} 
\end{array}
\right)\;.
\end{equation}
So we see that most exotics decouple at the linear level in the VEV fields $\phi_i$, one pair
of exotic triplets gets masses at order nine in the $\phi_i$ fields and another
one at order three. One may speculate that this leads to the presence of colour
triplets somewhat below the GUT scale, which may account for the fact that,
within the MSSM, the strong fine structure constant
$\alpha_3=g_3^2/(4\pi)$ turns out to be about $3\,\%$ smaller than $\alpha_1$
and $\alpha_2$ at $M_\mathrm{GUT}$. An important
feature of the $\Z4^R$ symmetry is that integrating out the triplets does not give rise to 
dimension 5 proton decay operators, as each triplet that couples to quarks and
leptons pairs up with a triplet that does not (\cf\ the similar discussion in
\cite{Babu:1993we}). In other words, the mass partner $\delta$ of a
triplet $\overline{\delta}$ that couples to $Q_\ell\, L_k$ (and therefore has $\Z4^R$
charge 0) cannot couple to $Q_i\,Q_j$ (\Figref{fig:NoDeltaExchange}).

The Yukawa couplings are
\begin{subequations}
\begin{eqnarray}
 Y_u & = & \overline{L}_{1}\,\left(
\begin{array}{ccc}
 1 & 0 & 0 \\
 0 & \phi^4 & \phi^4 \\
 0 & \phi^4 & \phi^4
\end{array}
\right) + 
\overline{L}_{2}\,\left(
\begin{array}{ccc}
 \phi^{12} & 0 & 0 \\
 0 & \phi^4 & \phi ^4 \\
 0 & \phi^4 & \phi ^4
\end{array}
\right)\;,
\\
Y_d & = &
L_{1}\,\left(
\begin{array}{ccc}
 0 & \phi ^{22} & \phi^{22} \\
 \phi^{22} & 0 & 0\\
 \phi^{22} & 0 & 0
\end{array}
\right) + 
L_{4}\,\left(
\begin{array}{ccc}
 0 & 1 & \phi^{12} \\
 1 & 0 & 0\\
 \phi^{12} & 0 & 0
\end{array}
\right) \\
& & + L_{5}\,\left(
\begin{array}{ccc}
 0 & \phi^{12} & 1 \\
 \phi^{12} & 0 & 0\\
 1 & 0 & 0
\end{array}
\right)~=~ Y_e^T\;.
\end{eqnarray}
\end{subequations}
As already mentioned, $Y_d$ and $Y_e$ have rank 2.

A crucial property of the string embedding is that, unlike in the bottom-up
approach, we have obtained an understanding of the origin of the $\Z4^R$
symmetry. In stringy language discrete $R$ symmetries originate from what is
called ``$H$-momentum conservation'' \cite{Dixon:1986qv}, which reflects
discrete rotational symmetries of compact space-time dimensions. In our orbifold
we have three $\mathbbm{T}^2/\Z2$ orbifold planes which can be rotated against
each other by $180^\circ$. Each rotational symmetry manifests itself as a
discrete order four $R$ symmetry in the effective field-theoretic description of
the model. The remnant $\Z4^R$ discussed above is a linear combination of such
symmetries and other discrete symmetries, either coming from the space group
selection rule or emerging from continuous \U1 symmetries through spontaneous
breaking (see the general discussion in \cite{Petersen:2009ip}). Note also that
such discrete $R$ symmetries can already appear anomalous at the orbifold
point~\cite{Araki:2008ek}. That is, no mixing with the so--called anomalous \U1
is required to obtain a $\Z{M}^R$ symmetry whose anomalies are canceled by the
GS mechanism.\footnote{The generality of the $\Z{4}^R$ symmetry connected with a
$\Z{2}$ orbifold suggests that  $\Z{4}^R$ invariant models may also be obtained
in other orbifold constructions such as  $\Z{6}$--II or $\Z{4} \times \Z{2}$ 
orbifolds of $\mathbbm{T}^6$.}

Perhaps the most important property of the model is that there is a
proportionality between the holomorphic mass term connecting $L_1$ and
$\overline{L}_1$ and the superpotential VEV. This relation can be derived in an
\SU6 orbifold GUT limit of the model, where it emerges due to gauge invariance
in extra dimensions \cite{Brummer:2010fr} (cf.\ also the field--theoretic
discussion in \cite{Hebecker:2008rk,Brummer:2009ug}). Let us comment that the
same \SU6 gauge symmetry also enforces the tree--level equality between the
gauge and top--Yukawa couplings \cite{Hosteins:2009xk}. We hence see that, at
least in this model, the superpotential VEV is both a measure for the gravitino
mass, as usual, and the $\mu$ term.

\section{Singlet extensions}
\label{sec:SingletExtensions}

In \Secref{sec:SU5constraints} we have shown that the requirement of
universality for the mixed gauge anomalies constrains the order $M$ of a
potential $\Z{M}^R$ symmetry to be a divisor of 24.  As we have seen, this
analysis carries over in an obvious way to singlet extensions of the MSSM, 
since additional SM singlet fields cannot change the constraints coming from the
mixed gauge anomalies. In such extensions the MSSM subsector still has to obey
the criteria derived in \Secref{sec:SU5constraints}. However, the extra
(singlet) fields can be subject to additional symmetries.

In what follows we concentrate on two simple singlet extensions in which one or
two singlet fields, respectively, are added. The first part of the discussion,
\Secref{sec:NMSSM}, is on the so--called NMSSM, in which the singlet couples to
the Higgs bilinear and there are cubic self--interactions. \Secref{sec:StrongCP}
is dedicated to a singlet extension of the MSSM which is capable of addressing
the strong CP problem.

\subsection{NMSSM}
\label{sec:NMSSM}

In the NMSSM, there is one additional singlet $\singlet$ with superpotential
\begin{equation}
\label{eq:NMSSMsuperpotential}
 \mathscr{W}~=~\mathscr{W}_\mathrm{MSSM}^{\mu=0} + \lambda\, \singlet\, \hu\, \hd 
 + \kappa\, \singlet^3 \;.
\end{equation}
Let us now consider what this implies for the order $M$ of a $\Z{M}^R$
symmetry.

\subsubsection{Constraints from NMSSM couplings}

There are three different classes of $\Z{M}^R$ symmetries for which the
$\singlet^3$--term of \Eqref{eq:NMSSMsuperpotential} implies different 
charges for the singlet $\singlet$, \ie
\begin{eqnarray}
\label{eq:0mod3}
 M&=& 0 \; \text{mod} \; 3 
 \quad \Rightarrow \quad \text{no $\singlet^3$ term possible}\;, \\
\label{eq:1mod3}
 M&=& 1 \; \text{mod} \; 3 \quad \Rightarrow \quad 
 q_\singlet ~=~ \frac{M+2}{3}\; \mod M \;, \\
\label{eq:2mod3}
 M&=& 2 \; \text{mod} \; 3 \quad \Rightarrow \quad 
 q_\singlet ~=~ \frac{2M+2}{3}\; \mod M \;,
\end{eqnarray}
with $q_\singlet$ the $\Z{M}^R$ charge of $\singlet$.

\paragraph{$\boldsymbol{M= 1 \; \text{mod} \; 3}$.}
Let us first consider the case $M= 1 \; \text{mod} \; 3$. The term $\lambda \singlet
\hu \hd$ together with \Eqref{eq:anomuniv1a} then implies
\begin{eqnarray}
 \left( \frac{M+2}{3} \; \text{mod}  \; M \right) 
 +  \left(4 \; \text{mod}  \; 2  \eta \right)
 &=& 2 \; \text{mod}  \; M \nonumber \\
\Rightarrow \frac{M+8}{3} &=& 0 \; \text{mod}  \; M \;.
\end{eqnarray}
This equation has only one non--trivial solution for integer $M$, namely $M=4$. 
Note that in this case $q_\singlet = 2 \mod 4$ and a linear term in  $\singlet$
is also allowed in the superpotential. Strictly speaking this is  not the NMSSM
but it is viable if the linear term is very small. We will  discuss later why
this may be natural.

Following the analysis of \Secref{sec:SU5constraints} and using
\Eqref{eq:1mod3}, the unique charge assignment compatible with the Weinberg
operator is  shown in \Tabref{tab:NMSSM1}.
\begin{table}[h]
\centerline{
\begin{tabular}{|c|c|c|c|c|c|}
\hline
 $M$ & $q_{\boldsymbol{10}}$ & $q_{\overline{\boldsymbol{5}}}$ & $\qhu$ & $\qhd$ & $q_\singlet$ \\
\hline
 4 & 1 & 1 & 0 & 0 & 2 \\
 \hline
\end{tabular}
}
\caption{Charge assignments for the $\Z{4}^R$ symmetry.}
\label{tab:NMSSM1}
\end{table}

This is exactly the $\Z{4}^R$ symmetry which we discussed in
\Secref{sec:SimpleZ4R}. We have seen that the mixed gauge anomaly
coefficients of this symmetry satisfy the Green--Schwarz condition. Of course the
singlet does not change these coefficients, so the analysis still applies.

\paragraph{$\boldsymbol{M= 2 \; \text{mod} \; 3}$.}
Let us now consider the case $M= 2 \; \text{mod} \; 3$.
The term $\lambda \singlet \hu \hd$ together with \Eqref{eq:anomuniv1a} then implies
\begin{eqnarray}
\frac{2M+8}{3} &=& 0 \; \text{mod}  \; M \;.
\end{eqnarray}
The solutions to this equation are $M=2,8$. As we have noted earlier there are
no meaningful $M=2$ $R$ symmetries. The $M=8$ case however is very interesting
since, in this case, $q_\singlet = 6 \mod 8$ and the linear term in $\singlet$
is forbidden.  Following the analysis of \Secref{sec:SU5constraints} and using
\Eqref{eq:2mod3}, the unique charge assignment compatible with the
Weinberg operator is shown in \Tabref{tab:NMSSM2}.

\begin{table}[h]
\centerline{%
\begin{tabular}{|c|c|c|c|c|c|}
\hline
 $M$ & $q_{\boldsymbol{10}}$ & $q_{\overline{\boldsymbol{5}}}$ & $\qhu$ & $\qhd$ & $q_\singlet$ \\
\hline
 8 & 1 & 5 & 0 & 4 & 6 \\
\hline
\end{tabular}
}
\caption{Charge assignments for the $\Z{8}^R$ symmetry.}
\label{tab:NMSSM2}
\end{table}

As the singlet does not contribute to mixed gauge anomalies, we know already
from \Tabref{tab:ChargeAssignments} that the $\mathbbm{Z}_8^R$ symmetry has
$A_{1\le i\le 3}^{R}(\mathrm{MSSM})=\rho=1$.

\subsubsection{The hierarchy problem}

Searching for possible $\Z{M}^R$ symmetries in the context of the NMSSM we found
that there are only two potential candidates: a $\mathbbm{Z}_4^R$  and a
$\mathbbm{Z}_8^R$ symmetry. The $\mathbbm{Z}_4^R$ symmetry is actually a
subgroup of the $\mathbbm{Z}_8^R$ symmetry, hence both symmetries are closely
related. While the $\mathbbm{Z}_4^R$ commutes with \SO{10} the $\mathbbm{Z}_8^R$
only commutes with \SU{5}. In both cases all dimension four and five baryon and
lepton number violating operators are forbidden (except for the Weinberg
operator), consistent with what we found in \Secref{sec:SU5constraints}. 

A potential problem with NMSSM models arises because SUSY breaking breaks the
$R$ symmetry and in radiative order a linear term in $N$ is generated in the
superpotential. If the coefficient of this linear term is larger than the square
of the electroweak scale it will lead to a large VEV for the  singlet $\singlet$
and therefore to a destabilization of the SUSY solution to the gauge hierarchy
problem.  This has been studied in detail by Abel~\cite{Abel:1996cr} who showed
that the only dangerous operators that induce divergent tadpoles arise either
from even terms in the superpotential or odd terms in the K\"ahler potential. He
also showed that an $R$ symmetry can avoid such terms  because of the different
$R$ charges of the super- and K\"ahler potential (\cf\ also
\cite{Panagiotakopoulos:1998yw}). From the charge assignments of
\Tabref{tab:NMSSM1} and \Tabref{tab:NMSSM2} for the singlet $\singlet$ and the
Higgs fields it is easy to show that the super- and K\"ahler- potentials
actually do have exactly this structure in both the $\Z4^R$ and the $\Z8^R$ case
and so in both cases radiative corrections do not destabilise the SUSY solution
to the hierarchy problem. 

The main difference between $\Z{4}^{R}$ and $\Z{8}^{R}$ is that the former
allows a linear term even at tree level. Does this mean that it is necessary to
have the full $\Z{8}^{R}$ symmetry when building the NMSSM? In effective
theories, such as those describing the massless degrees of freedom in string
compactifications, the superpotential starts with cubic terms in the fields and
the linear term only appears through the coupling of the singlet field to fields
acquiring VEVs. If the only (non--moduli) fields, $\phi$, with VEVs above the
electroweak scale are in the hidden  sector the coupling will be suppressed by
messenger field masses, $M_*$, which may be as large as the Planck scale.
Allowing for trilinear couplings to messenger fields as well as trilinear
couplings between messenger and hidden sector fields and assuming no additional
symmetries, the leading term in the superpotential after integrating out the
messenger fields is $N\phi^{4}/M_*^{2}$ with the messenger scale $M_*$. Taking
Planck scale messengers, the constraint that this should not disturb the
hierarchy is that $\langle \phi \rangle \le\sqrt{M_{W}M_\mathrm{P}}$ which is
satisfied if the dominant VEV comes from the SUSY breaking sector. In this case
it is sufficient to impose just the $\Z{4}^{R}$ symmetry when building the
NMSSM.

The role of the SM singlets $\psi_2^{(i)}$ with $R$--charge 2 (such as \singlet\
for the case $\Z4^R$) has recently been discussed in the context of singlet
(moduli) stabilization~\cite{Kappl:2010yu}. There it was found that for a
superpotential with generic coefficients the number of singlets with $R$--charge
2 should not exceed the number of fields $\phi_0^{(j)}$ with $R$--charge 0 since
otherwise the $F$--term conditions would overconstrain the system. Moreover, the
$\psi_2^{(i)}$ fields pair up with an equal number of $\phi_0^{(j)}$ fields.
That is, for generic superpotential coefficients one might not expect to find
vacua with an unbroken $\Z4^R$ symmetry and a massless singlet with $R$--charge
2. However, it is quite conceivable that there are symmetries between the
$F$--terms. In such a situation the $\psi_2^{(i)}-\phi_0^{(j)}$ mass matrix
won't have full rank such that one is effectively left with one (or more)
singlet(s) with $R$--charge 2. It will be interesting to see if this situation
can be realized in string models in which there are additional symmetries, such
as $D_4$ \cite{Kobayashi:2004ya,Kobayashi:2006wq}, relating the superpotential
coefficients.

\subsubsection{Non--perturbative effects}
\label{sec:Z8R}

Non--perturbative effects may also be important in determining the low energy
phenomenology. From \Eqref{eq:DeltaWnp} we see that the superpotential has a
term of the form
\begin{equation}
\Delta\mathscr{W}_\mathrm{np} ~=~B_{0}\,\mathrm{e}^{-b\,\dilaton}
\label{eq:np}
\end{equation}
with a constant $b$.  This parametrizes the non--perturbative effects discussed
above, and may be interpreted as a hidden sector gaugino
condensate. It provides the order parameter for local supersymmetry and
generates the gravitino mass 
\begin{equation}
\frac{ \langle\Delta\mathscr{W}_\mathrm{np}\rangle}{M_\mathrm{P}^{2}}~\sim~\frac{\langle \lambda\lambda\rangle}{M_\mathrm{P}^2}
 ~\sim~m_{3/2}\;.
\end{equation}
$\Delta\mathscr{W}_\mathrm{np}$ has $R$--charge 2 (\cf\ the discussion in
\Secref{sec:SimpleZ4R}) and similar non-perturbative effects can contribute to
further terms in the superpotential.  The crucial property of the
non-perturbative couplings is that they are naturally suppressed. To
parametrize these effects we denote by a superfield $Y$ a non--perturbative 
term of the form given in \Eqref{eq:np} (scaled by the factor
$M_\mathrm{P}^{-2}$) carrying $R$--charge 2 and we construct the superpotential
involving $Y$ that is consistent with the relevant $R$ symmetry.

The lowest superpotential terms in $Y$ have the form
\begin{eqnarray}
 \Delta \mathscr{W}_{\Z{4}^{R}}
 & = & Y + Y^2  \singlet + Y\, \singlet^2 + \,Y\,  \hu\, \hd \nonumber  \\
 & \sim & m_{3/2}\, M_\mathrm{P}^2 + m_{3/2}^2\, \singlet + m_{3/2}\, \singlet^2 
               + m_{3/2}\, \hu\, \hd \;,\\
 \Delta  \mathscr{W}_{\Z{8}^{R}}
 & = & Y + Y^2 \left( \singlet + Y\, \singlet^2 + \,Y\,  \hu\, \hd \right) \;\nonumber \\
 & \sim & 
 m_{3/2}\, M_\mathrm{P}^2 + m_{3/2}^2\, \singlet 
 + \frac{m_{3/2}^3}{M_\mathrm{P}^2}\, \singlet^2 
               + \frac{m_{3/2}^3}{M_\mathrm{P}^2}\, \hu\, \hd\;.\label{eq:WNMSSM1}
\end{eqnarray}
All of these terms have magnitude determined by the gravitino mass scale. For
gauge mediation this scale can be very small and these terms negligible. For
gravity mediation however the gravitino mass scale is the scale of supersymmetry
breaking in the visible sector and the unsuppressed  terms cannot be neglected.
In this case, the magnitude of the $\Delta \mathscr{W}_{\Z{4}^{R}}$ terms is
such as to reproduce the superpotential of the S--MSSM
\cite{Cassel:2009ps,Delgado:2010uj}, where, apart from the usual NMSSM couplings
also holomorphic mass terms for the singlets and the Higgs fields of the order
$m_{3/2}$ are introduced. This extension of the SM has been shown to
significantly reduce the fine tuning needed to accommodate the LEP Higgs mass
bound~\cite{Cassel:2009ps,Delgado:2010uj}. Our analysis yields a justification for the small holomorphic terms,
which have so far just been imposed by hand.
 
Interestingly the form of the non--perturbative effects is very sensitive to the
underlying symmetry. For the case of $\Z{4}^{R}$ there are additional
unsuppressed  linear and quadratic terms in $\singlet$ as well as a
non--perturbative contribution to the Higgsino mass.  For the case of
$\Z{8}^{R}$ only the linear term in $\singlet$  is unsuppressed. Because the
magnitude of all these terms is determined by the gravitino mass they will not
disturb the SUSY solution to the hierarchy problem. However,  for the case of
gravity mediation, the terms cannot be neglected and may be expected to
significantly change the NMSSM phenomenology. Given the different
non--perturbative  terms appearing in the $\Z{4}^{R}$ and $\Z{8}^{R}$ we may
expect these to have different phenomenological implications.

\subsection{Discrete $\boldsymbol{R}$ symmetries and the strong CP problem}
\label{sec:StrongCP}

The axion solution to the strong CP problem remains the most convincing to date.
Since it is based on the existence of a Peccei--Quinn (PQ) symmetry that forbids
the $\mu$ term it is of interest to ask whether a discrete $R$ symmetry can play
the role of the PQ symmetry.  Let us start by briefly discussing the role of
axions in our setup with $\Z{M}^{R}$ symmetries with particular focus on a
possible solution of the strong CP problem.  One potential candidate for such an
axion is the universal Green--Schwarz axion, $a=\im S$,
\cf~\Eqref{eq:dilatonaxion}. As discussed in \Apref{app:KaehlerStabilization},
in the case of K\"ahler stabilisation we are --- to leading order --- left with
a massless GS axion. However, as was e.g.\ shown in~\cite{Fox:2004kb}, the
corresponding axion decay constant is of order the Planck scale if we demand the
usual value for the unified gauge coupling. This is well outside the
cosmologically allowed range $10^{10}\,\mathrm{GeV} < f_a < 3\times
10^{11}\,\mathrm{GeV}$, assuming no fine tuning in the initial axion
VEV~\cite{Fox:2004kb}. 

Kim and Nilles \cite{Kim:1983dt} have proposed an interesting model that
naturally gives an axion decay constant in the favoured range. They achieve this
by requiring that the coupling of the MSSM singlet field that contains the axion
to the Higgs supermultiplets be quadratic with a superpotential of the form
\begin{equation}
\label{eq:Ch5mupert}
\mathscr{W} ~=~\frac{\alpha}{M_\mathrm{P}} \singlet^2\, \hu\hd \;.
\end{equation}
If the inverse mass scale, $M_\mathrm{P}$, associated with this operator (the
mediator scale) is taken to be the Planck scale with $\alpha=\mathcal{O}(1)$ an
electroweak scale $\mu$ term of $\mathcal{O}(m_{3/2})$ is generated if the singlet VEV
($\langle N \rangle \equiv f_a$) is in the desired range. This corresponds to
$\langle N \rangle= \mathcal{O}(\sqrt{m_{3/2}M_\mathrm{P}}$) for the case of gravity mediated supersymmetry
breaking with $m_{3/2}$ close to the electroweak breaking scale. Furthermore,
the theory has a global (accidental) $\U1$ PQ symmetry under
which $\hu\hd$ and the singlet $\singlet$  transform non--trivially. Hence, the
VEV of $\singlet$ breaks the PQ symmetry and for  the case of gravity mediated
supersymmetry breaking the scale of breaking is of $\mathcal{O}(10^{10}-10^{11}
\gev)$. So the associated axion coming from the singlet $\singlet$ has the right
properties to solve the strong CP--problem~\cite{Kim:1983dt}.

One has to ensure that higher--dimensional operators, which explicitly break 
the PQ symmetry, do not spoil the solution.\footnote{A continuous $R$ symmetry
can protect the PQ symmetry at higher orders if it is broken only by the
superpotential VEV and the intermediate scale VEVs of singlets
\cite{Choi:2010xf}.} An additional complication in the construction  of a viable
model is that one needs an $|\singlet|^{6}$ term in the scalar potential to get 
an intermediate scale VEV of the correct magnitude if the soft supersymmetry
breaking mass  of the $\singlet$ field is $\mathcal{O}(m_{3/2})$. Since a
superpotential term $\singlet^4$  breaks the PQ symmetry too strongly to give a
viable axion at least one additional  singlet (called $X$) is  needed to
generate the $|\singlet|^6$ term in the $F$--term potential. Let us consider
whether the $\Z{M}^{R}$ symmetries can give such a structure.


In order to construct a viable model with intermediate scale breaking we start
with the superpotential
\begin{equation}
\label{eq:mupert1}
 \mathscr{W} 
 ~=~
 \frac{\alpha}{M_\mathrm{P}} \singlet^2\, \hu\, \hd +\frac{\beta}{M_\mathrm{P}}X \singlet^{3}\;.
\end{equation}
which requires $\Z{M}^{R}$ charges $q_\singlet=-1$ and $q_X=5$,
\cf~\Tabref{tab:ChargeAssignments}. Including soft SUSY breaking mass terms,
this gives the potential
\begin{eqnarray}
 \mathscr{V} 
 & = &
 \left|\frac{2\alpha}{M_\mathrm{P}} \singlet\,\hu\,\hd
 +\frac{3\beta}{M_\mathrm{P}}X\, \singlet^{2}\right|^{2}
 +\left|\frac{\beta}{M_\mathrm{P}} \singlet^{3}\right|^{2} 
 +m_{\singlet}^{2}\,|\singlet|^{2}+m_{X}^{2}\,|X|^{2}\;
 \nonumber \\
  & & \label{eq:mupert2}
\end{eqnarray}
that has the required $|\singlet|^{6}$ stabilising term.
Provided $m_{X}^2$ is negative, the field $X$ acquires
a VEV 
\begin{equation}
\label{eq:mupert}
\langle X \rangle~\sim~
\mathcal{O}(\langle \singlet\rangle^{3}/M_\mathrm{P}^{2} )\;,
\end{equation}
where we have used $\langle\hu\rangle\sim \langle\hd\rangle\sim
(\langle\singlet\rangle^{2}/M_\mathrm{P}) =\mathcal{O}(m_{3/2})$. The
superpotential in \Eqref{eq:mupert1} has a global $\U1$ PQ symmetry under which 
$\hu\hd$, $\singlet$ and $X$ transform non--trivially. However the
superpotential is not the most general one allowed by an underlying discrete
$R$ symmetry  for this will allow additional terms of the form
$\singlet^{p}X^{q}$ for $p,q$ integer, where the values of $p,q$ are constrained
by the choice of $\Z{M}^R$.  Such terms will break the PQ symmetry generating a
mass for the  would--be axion. If the axion solution to the strong CP problem is
to be maintained this contribution to the mass should be five orders of
magnitude smaller than the corresponding contribution from QCD, $\delta m_{a}
\lesssim 10^{-5} m_a^\mathrm{QCD} \lesssim
10^{-9}\ev$~\cite{Kamionkowski:1992mf,Barr:1992qq}.  This puts a strong
constraint on the discrete symmetry for it must be large enough to forbid the
additional terms to a high order. 

Including the $\singlet^{p}X^{q}$ term in \Eqref{eq:mupert1} one sees that it is
the interference between the last two terms in $|F_{X}|^{2}$ that gives the 
dominant contribution because the VEV of $X$ is smaller than the VEV of
$\singlet$. This term is of
$\mathcal{O}(\singlet^{p+3}X^{q-1}/M_\mathrm{P}^{p+q+1})$ in the potential  and
gives a contribution to the axion mass given by  
\begin{equation}
\delta m_{a}^{2}~=~\mathcal{O}(\singlet^{p+1}X^{q-1}/M_\mathrm{P}^{p+q-2})=\mathcal{O}(\singlet^{p+3q-2}/M_\mathrm{P}^{p+3q-4}) \;.
\end{equation}
The leading terms for the candidate $\Z{M}^{R}$ symmetries (for which $5q-p-2= 0
\mod M$)  are $\singlet^2,\; \singlet^{4},\;X^{2},\; \singlet X^{3}$ and
$\singlet^8 X^{2}$ corresponding to $p+3q-2=0,\;2,\;4,\;8$ and $12$  for
$M=4,\;6,\;8,\;12$ and $24$ respectively. The constraint $\delta
m_{a}^{2}<10^{-18}\ev^{2}$ is equivalent to $p+3q-2>8$ so we see that only
$\Z{24}^{R}$  is large enough to accommodate this method of simultaneously
generating the $\mu$ term and solving the strong CP problem.

Note that the singlets also induce baryon and lepton number violation as well as
a small amount of $R$ parity violation.  In the case of the $\Z{24}^R$ the
leading order terms (\cf~\Tabref{tab:ChargeAssignments}) are given by 
\begin{equation}
\frac{X \singlet^{3}}{M_\mathrm{P}^5} \rep{10}\,\rep{10}\,\rep{10}\, \crep{5}  
\quad\text{and}\quad 
\frac{X \singlet^{2}}{M_\mathrm{P}^3}
\rep{10} \,\crep{5}\, \crep{5} \;.
\end{equation}
With intermediate scale VEVs for $\singlet$ and $\langle X\rangle
=\mathcal{O}(\langle\singlet\rangle^{3}/M_\mathrm{P}^{2})$ the contribution  of
these operators to nucleon decay is strongly suppressed lying below the
irreducible dimension 6 operator contribution. Also the $R$ parity violation is
negligible.  In summary, a singlet extension of the MSSM with a $\Z{24}^R$
symmetry provides us with a simultaneous solution to the $\mu$ and strong CP
problems. The phenomenological implications for the Higgs structure may
differ significantly from the NMSSM and remain to be analysed.

\section{Conclusions}
\label{sec:Conclusions}

We have discussed possible discrete symmetries for the MSSM which commute with
\SU5. We have seen that, in order to address the $\mu$ problem, these have to be
$R$ symmetries. We have surveyed all possible discrete $\Z{M}^R$ symmetries.
Anomaly cancellation requires that the order $M$ be a divisor of $24$. We
identified 5 phenomenologically viable symmetries for the MSSM. 

The simplest of the 5 MSSM symmetries is a $\Z4^R$ which commutes with \SO{10}. 
This symmetry forbids all $R$--parity violating couplings, dimension five proton
decay operators and the $\mu$ term at tree--level while allowing the usual
Yukawa couplings and the neutrino mass operator. At the non--perturbative level
the $\mu$ term and the dimension five proton decay operators are generated. We
argued that in settings in which supersymmetry breaking is related to some
non--perturbative dynamics the $\mu$ term will be of the order of the MSSM soft
terms. In particular, in gravity mediation we will have $\mu\sim m_{3/2}$ and
coefficients of the dimension five proton decay operators
$\kappa^{(1,2)}_{ijk\ell}\sim m_{3/2}/M_\mathrm{P}^2$, \ie\ sufficiently
suppressed. Thus the $\Z4^R$ symmetry provides us with a simultaneous solution
to the arguably two most severe problems of the MSSM.
We have discussed how to embed the $\Z4^R$ into string theory. Specifically, we
have constructed a $\Z2\times\Z2$ orbifold with this $\Z4^R$ and the exact MSSM
spectrum below the compactification scale, in which the $\Z4^R$ originates from
the Lorentz symmetry of compactified dimensions. At the non--perturbative level
the $\mu$ term is, due to the $\Z4^R$ anomaly, generated. There is an exact
matter parity and dimension five proton decay is well below experimental
limits. 

We have discussed the role of discrete symmetries in singlet extensions of the
MSSM. There are two possible symmetries consistent with the structure of the
NMSSM, $\Z4^R$ and $\Z8^R$, both of which are capable of solving the hierarchy
problem. The $\Z8^R$ allows the usual couplings while forbidding the linear term
for the singlet at the perturbative level. In the $\Z4^R$ case, one obtains
holomorphic mass terms for the singlet and the Higgs at the non--perturbative
level. We have argued that the size of such terms is of the order $m_{3/2}$,
leading to an S--MSSM--like scheme in which the smallness of the explicit mass
terms for the singlets and Higgs finds an explanation. As another application we
have discussed how discrete $R$ symmetries can lead to approximate PQ \U1
symmetries capable of solving the strong CP problem. Given the upper bound on
the order of $\Z{M}^R$,  $M\le 24$, we identified $\Z{24}^R$ as the unique
possibility for solving the $\mu$- and strong CP problems simultaneously.


\subsection*{Acknowledgments}

We would like to thank Oleg Lebedev for useful discussions. This research was
supported by the DFG cluster of excellence Origin and Structure of the Universe,
the \mbox{SFB--Transregio} 27 ``Neutrinos and Beyond'', LMUExcellent and the
Graduiertenkolleg ``Particle Physics at the Energy Frontier of New Phenomena''
by Deutsche Forschungsgemeinschaft (DFG).  S.R.\ acknowledges partial support
from DOE grant DOE/ER/01545-892. The research presented here was partially
supported by the EU ITN grant UNILHC 237920 (Unification in the LHC era) and by
the UK Science and Technology Facilities Council Oxford particle theory rolling
grant ST/G000492/1. H.M.L.\ is supported by the Korean-CERN fellowship.


\appendix

\section{Discrete anomalies in the path integral approach}
\label{app:PathIntegral}

In this appendix we re--derive Abelian discrete anomalies with the path integral
method, following~\cite{Araki:2006mw,Araki:2008ek}. Among other things, we will
describe how this allows us to understand the discrete version of the
Green--Schwarz mechanism.

\subsection{Path integral derivation of anomalies}

Consider a theory described by a \Lagrangean\ $\mathscr{L}$ with a set of
fermions $\Psi=[\psi^{(1)},\dots,\psi^{(M)}]$,
where $\psi^{(m)}$ denotes a field transforming in the irreducible representation (irrep)
$\boldsymbol{R}^{(m)}$ of all internal symmetries.
A general transformation $\Psi~\to~U\,\Psi$
or, more explicitly,
\begin{equation}\label{eq:Trafo2}
 \left[\begin{array}{c}
 \psi^{(1)}\\ \vdots\\ \psi^{(M)}\end{array}\right]
 ~\to~
 \left(\begin{array}{ccc}
 U^{(1)} &  & 0\\
 & \ddots & \\
 0 & & U^{(M)}
 \end{array}\right)\,
 \left[\begin{array}{c}
 \psi^{(1)}\\ \vdots\\ \psi^{(M)}\end{array}\right]\;,
\end{equation}
which leaves $\mathscr{L}$ invariant (up to a total derivative) denotes a
classical symmetry.

A classical symmetry implies that certain correlators  vanish at the classical
level. To see this, consider the correlator
\begin{equation}
 C_{n_1\dots n_M}~=~\left\langle(\psi^{(1)})^{n_1}\cdots(\psi^{(M)})^{n_M}\right\rangle\;.
\end{equation}
Now, if the field combination $(\psi^{(1)})^{n_1}\cdots(\psi^{(M)})^{n_M}$ is not
invariant under the symmetry transformation, we arrive at the (premature)
conclusion that $C_{n_1\dots n_M}=0$.

Classical chiral symmetries  can be broken by quantum effects, \ie\ have an
anomaly. Specifically, consider a chiral transformation
\begin{equation}\label{eq:ChiralTransformation}
 \Psi(x) ~ \to ~ \Psi'(x)~=~\exp\big(\I\,\alpha\,P_\mathrm{L}\big)\,\Psi(x)\;,
\end{equation}
where $\alpha=\alpha^\mathrm{anom}\mathsf{T}_\mathrm{anom}$ with
$\mathsf{T}_\mathrm{anom}$ denoting the generator of the transformation and
$\alpha^\mathrm{anom}$ being a parameter, and $P_\mathrm{L}$ is the
left--chiral projector.

We wish now to show that this implies vanishing correlators at the
classical level may appear at the quantum level. To this end, write the
correlator as a path integral,
\begin{equation}
 C_{n_1\dots n_M}
 ~=~
 \int\!\mathcal{D}\Psi\,\mathcal{D}\overline{\Psi}\,
 (\psi^{(1)})^{n_1}\cdots(\psi^{(M)})^{n_M}
 \,\mathrm{e}^{\I\,\mathcal{S}}\;,
\end{equation}
where $\mathcal{S}$ denotes the action, which is left invariant
under \eqref{eq:ChiralTransformation}.
Now recall that under the transformation \eqref{eq:Trafo2} the path integral
measure undergoes a non--trivial change \cite{Fujikawa:1979ay,Fujikawa:1980eg},
\begin{equation}
 \mathcal{D}\Psi\,\mathcal{D}\overline{\Psi}
 ~\to~J(\alpha)\,
 \mathcal{D}\Psi\,\mathcal{D}\overline{\Psi}\;,
\end{equation}
where the Jacobian of the transformation is given by
\begin{equation}\label{eq:Jacobian}
 J(\alpha)
 ~=~\exp\left\{\I\,\int\!\D^4x\,\mathcal{A}(\alpha)\right\}\;.
\end{equation}
The crucial observation is now that in the presence of a non--trivial Jacobian
the full quantum correlator can be invariant. This is true regardless of
whether the transformation \eqref{eq:Trafo2} is continuous or discrete, or
whether it is gauged or global.

The anomaly function $\mathcal{A}$ appearing in \eqref{eq:Jacobian} decomposes
into a gauge and a gravitational part
\cite{AlvarezGaume:1983ig,AlvarezGaume:1984dr,Fujikawa:1986hk},
\begin{equation}\label{eq:AnomalyFunctionA}
 \mathcal{A}(\alpha)~=~
  \mathcal{A}_\mathrm{gauge}(\alpha)+\mathcal{A}_\mathrm{grav}(\alpha)
\;,
\end{equation}
with
\begin{eqnarray}
 \mathcal{A}_\mathrm{gauge}(\alpha) & = & \frac{1}{32\,\pi^2}
 \Tr\left[\alpha\,\mathcal{F}\widetilde{\mathcal{F}}\right]\;,
\label{eq:Agauge} \\
\mathcal{A}_\mathrm{grav}(\alpha) & = & 
-\frac{1}{384\,\pi^2}  \mathcal{R} \widetilde{\mathcal{R}} \, \Tr \left[\alpha \right] 
\;.
\end{eqnarray}
We have suppressed index contractions, \ie\ 
$\mathcal{F}\widetilde{\mathcal{F}} = \mathcal{F}^{\mu\nu}\,\widetilde{\mathcal{F}}_{\mu\nu}$.
Here $\mathcal{F}_{\mu\nu}=[D_\mu,D_\nu]$ is the field strength of the gauge
symmetry, such that $\mathcal{F}_{\mu\nu}=(\partial_\mu A_\nu-\partial_\nu
A_\mu)$ for a \U1 symmetry, $\mathcal{F}_{\mu\nu}=F_{\mu\nu}^a\,\mathsf{T}_a$
for non--Abelian gauge groups, and
$\widetilde{\mathcal{F}}^{\mu\nu}=\frac{1}{2}\varepsilon^{\mu\nu\rho\sigma}\mathcal{F}_{\rho\sigma}$
denotes its dual. Similarly, $\mathcal{R}$ represents the Riemann curvature tensor and $\mathcal{R} \widetilde{\mathcal{R}}=\frac{1}{2}\varepsilon^{\mu\nu\rho\sigma} \mathcal{R}_{\mu\nu}^{\phantom{\mu\nu}\lambda\gamma} \mathcal R_{\rho\sigma\lambda\gamma} $ . The trace `Tr' runs over all internal indices.

Now we specialize to the case where $\alpha$ is a \Z{M} transformation. For the
anomaly to be absent, \ie\ $J(\alpha)=1$, we arrive at the 
conditions~\cite{Araki:2006mw,Araki:2008ek}
\begin{subequations}\label{eq:PathIntegralConstraints}
\begin{eqnarray}
 G-G-\Z{M}:&& \quad \sum_{\boldsymbol{r}^{(f)}} 
 	\ell\left(\boldsymbol{r}^{(f)}\right) \,
 	q^{(f)}~=~0 \mod \eta \;, \\
 \text{grav}-\text{grav}-\Z{M}:&& \sum_m q^{(m)}~=~0\mod \eta \;,
\end{eqnarray}
\end{subequations}
where (\cf\ equation~\eqref{eq:eta})
\[
 \eta~=~\left\{\begin{array}{ll}
    M & \text{for $M$ odd}\;,\\
    M/2 & \text{for $M$ even}\;
 \end{array}\right.
\]
and $q^{(m)}$ denotes the \Z{M} charge. The first sum runs over all irreducible
representations $\boldsymbol{r}^{(f)}$ of $G$ with Dynkin index
$\ell(\boldsymbol{r}^{(f)})$ while the second sum runs over all fermions.  Our
conventions are such that $\ell(\boldsymbol{N})=1/2$ for \SU{N} and
$\ell(\boldsymbol{N})=1$ for \SO{N}.
\eqref{eq:PathIntegralConstraints} are the traditional discrete anomaly
conditions~\cite{Ibanez:1991hv,Ibanez:1991pr} with the difference that the
$\Z{M}^3$ constraints do not appear; we will discuss $\Z{M}^3$ anomalies
separately in \Secref{sec:ZN3}.

\subsection{Green--Schwarz mechanism and re--derivation of $\boldsymbol{\delta_\mathrm{GS}}$}
\label{app:GS}

Consider a theory with simple gauge group $G$ and an `anomalous' Abelian gauge
factor $\U1_\mathrm{anom}$. Under $\U1_\mathrm{anom}$ the dilaton superfield $S$
shifts according to
\begin{equation}\label{eq:TrafoDilaton}
 \dilaton~\to~\dilaton+\frac{\I}{2}\delta_\mathrm{GS}\,\Lambda(x)
\end{equation}
with $\Lambda$ denoting the $\U1_\mathrm{anom}$ transformation, \ie\ the chiral
superfields follow the rule
\begin{equation}
 \Phi^{(f)}~\to~\mathrm{e}^{-\I\,Q_\mathrm{anom}^{(f)}\,\Lambda}\,\Phi^{(f)}\;.
\end{equation}
The corresponding transformation of the vector superfield $V_\mathrm{anom}$ is
\begin{equation}\label{eq:TrafoVanom}
 V_\mathrm{anom}~\to~V_\mathrm{anom}+\frac{\I}{2}(\Lambda-\Lambda^\dagger)
\end{equation}
with $\re \Lambda|_{\theta=0}=\alpha$.
In what follows, we derive the Green--Schwarz coefficient $\delta_\mathrm{GS}$ from the
requirement of invariance of the full action.

The dilaton--dependent part of the \Lagrangean\ is
\begin{eqnarray}
 \mathscr{L}_\mathrm{dilaton}
 & = &
 -\int\!\D^4\theta\,\ln\left(\dilaton+\dilaton^\dagger-\delta_\mathrm{GS}\,V_\mathrm{anom}\right)
 \nonumber\\
 & & {}+\left[
 \int\!\D^2\theta\,\frac{\dilaton}{4}\Tr W_\alpha W^\alpha+\text{h.c.}
 \right]
 \nonumber\\
 &&{}+\text{gravity terms}\;.
\end{eqnarray}
The first line of this \Lagrangean\ is already invariant under the combined
transformations \eqref{eq:TrafoDilaton} and \eqref{eq:TrafoVanom}. The trace in
the second line is supposed to run over all gauge factors, including
$\U1_\mathrm{anom}$.

Decomposing the scalar component of the dilaton into a real and an imaginary
(or axionic) part,
\begin{equation}
 \dilaton|_{\theta=0}~=~s+\I\,a\;,
\end{equation}
leads to the usual couplings of the axion $a$
\begin{equation}\label{eq:BosonicGaugeTerms}
 \mathscr{L}
 ~\supset~
 -\frac{a}{8} F_\mathrm{anom} \widetilde{F}_\mathrm{anom} -\frac{a}{8} F^{a} \widetilde{F}^a  +
 \frac{a}{4}  \mathcal{R}\widetilde{\mathcal R}\;,
\end{equation}
where $F$ and $F_\mathrm{anom}$ denote the gauge field
strength of $G$ and $\U1_\mathrm{anom}$ respectively. 

Hence, under a $\U1_\mathrm{anom}$ transformation with parameter $\alpha$ the
axionic \Lagrangean\ shifts by
\begin{equation}
  \Delta\mathscr{L}_{\mathrm{axion}}  ~=~  
  -\frac{\alpha}{16}\delta_\mathrm{GS}\, F_\mathrm{anom} \widetilde{F}_\mathrm{anom} 
  -\frac{\alpha}{16}\delta_\mathrm{GS}\, F^{a} \widetilde{F}^a  
  + \frac{\alpha}{8}\delta_\mathrm{GS}\,  \mathcal{R}\widetilde{\mathcal R}
  \;.
\end{equation}
The Green--Schwarz term $\delta_\mathrm{GS}$ can now be inferred by demanding
that the transformation of the axion $a$ cancels the anomalous variation of the
path integral measure. The latter can be absorbed in a change of the
\Lagrangean\ 
\begin{eqnarray}
 \Delta\mathscr{L}_{\text{anomaly}} &= &  \frac{\alpha}{32\pi^2} F_\mathrm{anom} \widetilde{F}_\mathrm{anom}\, \,A_{\U1_\mathrm{anom}^3} \nonumber \\
&&{} + \frac{\alpha}{32\pi^2} F^{a} \widetilde{F}^a  \, A_{G-G-\U1_\mathrm{anom}} \nonumber \\
&&{} - \frac{\alpha}{384\pi^2}  \mathcal{R}\widetilde{\mathcal R}  \, A_{\text{grav}-\text{grav}-\U1_\mathrm{anom}}
\;.
\end{eqnarray}
The coefficients $A$ are the anomaly coefficients given by
\begin{subequations}
\label{eq:anomaly_coefficients}
\begin{eqnarray}
 A_{\U1_\mathrm{anom}^3} & = &\frac{1}{3}\sum_m (Q_\mathrm{anom}^{(m)})^3 ~=~\frac{1}{3} \tr
 Q_\mathrm{anom}^3\;, \\
 A_{\text{grav}-\text{grav}-\U1_\mathrm{anom}} & = &\sum_m Q_\mathrm{anom}^{(m)} ~=~ \tr Q_\mathrm{anom}\;, \\
 A_{G-G-\U1_\mathrm{anom}} & = &\sum_{\boldsymbol{r}^{(f)}} \ell(\boldsymbol{r}^{(f)}) \, Q_\mathrm{anom}^{(f)} \:,
\end{eqnarray} 
\end{subequations}
where $Q_\mathrm{anom}^{(m)}$ denotes the $\U1_{\mathrm{anom}}$ charge. The
first two sums run over all left--handed Weyl fermions while the last sum
runs over all irreducible representations $\boldsymbol{r}^{(f)}$ of $G$ and
$\ell(\boldsymbol{r}^{(f)})$ is the Dynkin index.

The axion shift allows us to cancel the grav--grav--$\U1_{\mathrm{anom}}$,
$\U1_{\mathrm{anom}}^3$ and $G-G-\U1_{\mathrm{anom}}$ anomalies by demanding
$\Delta\mathscr{L}_{\mathrm{anomaly}} + \Delta\mathscr{L}_{\mathrm{axion}}=0$.
This fixes the Green--Schwarz constant to be given by
\begin{equation}\label{eq:AnomalyUniversalityU(1)}
 2\pi^2\,\delta_{\mathrm{GS}}
 ~=~
 \frac{1}{24}\tr Q_\mathrm{anom}
 ~=~
 \frac{1}{3}\tr Q_\mathrm{anom}^3 
 ~=~
 A_{G-G-\U1_\mathrm{anom}} \;,
\end{equation}
which is in agreement with the result obtained in a string
computation~\cite{Lerche:1987sg}.

\subsection{Discrete Green--Schwarz mechanism}
\label{app:DiscreteGS}

The Green--Schwarz mechanism also works if we replace $\U1_{\mathrm{anom}}$ by a
discrete \Z{M}. In this case the parameter $\alpha$ is no longer continuous but
$\alpha=\frac{2\pi n}{M}$ with some integer $n$. Of course, there is no gauge
field associated with the \Z{M}, \ie\ \Eqref{eq:TrafoVanom} does not apply here.
The discussion then goes as in the previous subsection. The discrete
Green--Schwarz constant is now defined in such a way that under the $\Z{M}$
transformation of fields
\begin{equation}
 \Phi^{(f)}~\to~\mathrm{e}^{-\I\,\frac{2\pi}{M}\,q^{(f)}}\,\Phi^{(f)}
\end{equation}
the dilaton shifts according to
\begin{equation}\label{eq:DiscreteShiftDilaton}
 \dilaton~\to~\dilaton+\frac{\I}{2}\Delta_\mathrm{GS}\;,
\end{equation}
where $\Delta_\mathrm{GS}$ is fixed only modulo $\eta$,
\begin{equation}\label{eq:DeltaGS}
 \pi\,M\,\Delta_\mathrm{GS}
 ~ \equiv ~
 \frac{1}{24}\,
 A_{\text{grav}-\text{grav}-\Z{M}} 
 ~ = ~
 A_{G-G-\Z{M}} 
 \mod \eta
 \;.
\end{equation}
The anomaly coefficients can be obtained from \Eqref{eq:anomaly_coefficients} by
replacing the $\U1_{\mathrm{anom}}$ charges $Q_\mathrm{anom}^{(m)}$ by the \Z{M}
charges $q^{(m)}$. Note that, unlike in the continuous case, the transformation
of the axion is only fixed modulo $\eta$. In the main body of the paper we
obtain constraints on possible discrete symmetries and charge assignments from
the requirement that \Eqref{eq:DeltaGS} possesses a solution, \ie\ that
the $A_{G-G-\Z{M}}$ coefficients for different gauge factors $G$ coincide modulo
$\eta$.

\subsection{A comment on $\boldsymbol{\Z{M}^3}$ anomalies}
\label{sec:ZN3}

If the discrete symmetry is embedded in a continuous symmetry, the universality
relations \eqref{eq:AnomalyUniversalityU(1)} also imply that
\begin{equation}
 A_{\mathrm{grav}-\mathrm{grav}-\Z{M}}
 -
 8\,A_{\Z{M}^3} ~=~0\mod \eta\;.
\end{equation}
with
\begin{equation}\label{eq:A-Z_N-cubed}
 A_{\Z{M}^3} ~=~ \mathcal{N}\,\sum_m (q^{(m)})^3 \;,
\end{equation}
where $\mathcal{N}$ is a normalization factor compensating for the rescaling of
the original \U1 charges $Q^{(m)}$ to integer \Z{M} discrete charges $q^{(m)}$. 
However, this relation is an embedding constraint rather than a true anomaly
constraint. That is, if this relation is not satisfied, this does not
necessarily imply a non--trivial variation of the path integral measure
\cite{Araki:2008ek}, and therefore it does not mean that classically forbidden
correlators will appear at the quantum level. 

The $\Z{M}^3$ anomaly constraints have lead to some confusion in the literature.
Banks and Dine \cite{Banks:1991xj} gave an argument for why there are no
$\Z{M}^3$ anomaly constraints. Following \cite{Ibanez:1991hv} they embedded the
\Z{M} into a \U1 symmetry; however they broke this down to a \Z{P\cdot M}
symmetry and were able to show that, while there is only a
\Z{M} for the chiral states, the extra heavy states `see' a \Z{P\cdot M} and can
be chosen such that the anomaly conditions following from the \U1 constraints
can be satisfied through their extra contributions.  

From this one might conclude that, in order to satisfy the $\Z{M}^3$ anomaly,
the true symmetry has to be $\Z{P\cdot M}$ rather than $\Z{M}$. However, this is
not necessarily the case as there can be  discrete symmetries that cannot be
obtained from continuous symmetries by \emph{spontaneous} breaking in four
dimensions. That is, the constraints from embedding \Z{M} in non--anomalous
continuous symmetries are sufficient to ensure anomaly freedom but not
necessary.

In order to be specific, let us look at the stringy origin of the $\Z4^R$
symmetry discussed in Section~\ref{sec:SimpleZ4R} of the main body of the paper.
The $\Z4^R$ has a clear geometric interpretation in terms of remnants of the
Lorentz group of compactified dimensions. However, some of the states
transforming non--trivially under our $\Z{4}^R$ are twisted states. These states
are chiral massless states which appear only \emph{after} orbifolding, \ie\
after we have broken the continuous Lorentz symmetry down to a discrete
subgroup. So it appears that there is no continuous interpolation between the
continuous \U1 and the discrete $\Z{4}^R$. Hence the derivation of discrete
anomalies based on embedding discrete symmetries in continuous symmetries might
not apply. On the other hand, the path integral method still works. This is
consistent with the fact that in our orbifold construction there is no
underlying $\Z{8}^R$ while we believe that the theory is UV complete.

\section{$\boldsymbol{\Z{M}}$ and $\boldsymbol{\Z{M}^R}$ anomaly coefficients}
\label{app:AnomalyCoefficients}

We start by looking at the MSSM amended by ordinary, \ie\ non--$R$, discrete
symmetries, where the fermions have the same charges as the superfields
$\Phi^{(f)}$ and turn then to the discussion of discrete $R$ symmetries.

\subsection{Anomaly coefficients for non--$\boldsymbol{R}$ $\boldsymbol{\Z{M}}$}
\label{B1}

The anomaly coefficients for discrete non--$R$ $\Z{M}$ symmetries are well
known~\cite{Ibanez:1991hv,Banks:1991xj,Ibanez:1991pr},
\begin{subequations}
\begin{eqnarray}
 A_{G-G-\Z{M}}& = &
 \sum_{\boldsymbol{r}^{(f)}} \ell(\boldsymbol{r}^{(f)})\cdot q^{(f)} 
 \;,\label{eq:A_G-G-ZN}\\
 A_{\mathrm{grav}-\mathrm{grav}-\Z{M}}
 & = & 
 \sum_m q^{(m)}
 \;.\label{eq:A_grav-grav-ZN}
\end{eqnarray}
These coefficients can be re--derived in the path integral approach
\cite{Araki:2008ek} (\cf\ \Apref{app:PathIntegral}).  In \Eqref{eq:A_G-G-ZN} we
sum over all irreducible representations $\boldsymbol{r}^{(f)}$ of $G$ while in
\Eqref{eq:A_grav-grav-ZN} we sum over all fermions. $\ell(\boldsymbol{r}^{(f)})$
denotes the Dynkin index of the representation $\boldsymbol{r}^{(f)}$. The
discrete charges $q$ are integers which are defined modulo $M$. Moreover, there
might be mixed \U1 anomalies if the normalization of the \U1 factors is known.
The coefficients are
\begin{equation}
 A_{\U1-\U1-\Z{M}}~=~
 \sum_m q^{(m)}\cdot \left(Q^{(m)}\right)^2
 \label{eq:A_U1-U1-ZN}
\end{equation}
\end{subequations}
with $Q^{(m)}$ denoting the normalized $\U1$ charges. We will discuss this
coefficient in more detail below.

Traditional anomaly freedom requires that for all anomaly coefficients
\begin{equation}\label{eq:anomaly-freedom}
 A~=~0\mod\eta \,.
\end{equation}
However, discrete anomalies can be canceled by the Green--Schwarz mechanism, in
which case one has to demand
\begin{eqnarray}
 A_{G-G-\Z{M}}
 & = &
 A_{\U1-\U1-\Z{M}}~= \frac{1}{24}~A_{\mathrm{grav}-\mathrm{grav}-\Z{M}}
 \nonumber\\
 & = & \rho\mod
 \eta
 \label{eq:anomaly-universality}
\end{eqnarray}

An important comment concerns the mixed \U1--\Z{M} anomaly coefficient
\eqref{eq:A_U1-U1-ZN}. Mixed
$\U1-\U1-\Z{M}$ anomalies are mostly ignored as they do not give meaningful constraints
unless one knows the normalization of the
charges~\cite{Ibanez:1992ji,Dreiner:2005rd}. 
Typically the sum in \Eqref{eq:A_U1-U1-ZN} is not invariant under shifting some discrete charges by
$M$. To see this, let us consider the example of hypercharge. We will denote the unnormalized $\U1_Y$ charge by $Q_Y^{(m)}$. 
The anomaly coefficient reads
\begin{equation}
 A_1 ~=~ \sum_m \frac{3}{5}\left(Q_Y^{(m)}\right)^2 q^{(m)}=\rho \mod \eta \:.
\end{equation}
We have the freedom to shift the \Z{M} charges by integer multiples of $M$, \ie\
we can define new \Z{M} charges $q'^{(m)}=q^{(m)}+k^{(m)} M$ with
$k^{(m)}\in\Z{}$. With the new charges the condition for anomaly freedom is
\begin{eqnarray}
 \quad \frac{3}{5}\sum_m \left( Q_Y^{(m)}\right)^2\,
 	\left(q^{(m)}+k^{(m)}\,M\right)  & = & \rho \mod \eta \\
 \Rightarrow  \quad A_1 + \frac{3}{5} M \underbrace{\sum_m k^{(m)}\left(
 Q_Y^{(m)}\right)^2}_{=:n}  &=&\rho \mod \eta \;.
\end{eqnarray}
We can choose the $k^{(m)}$ such that $n$ is an arbitrary integer because, for
example, $Q_Y(\overline{E})=1$. Hence, we arrive at
\begin{equation}
  A_1 ~=~\rho - \frac{3}{5}n\, M + m\, \eta \;, \qquad m\in\mathbbm{Z} \; .
\end{equation}
This can be rewritten as
\begin{eqnarray}
 M~\text{odd}~: \quad 5A_1 &=& 5\rho + (5m-3n)\,M\;, \\
 M~\text{even}~: \quad 5A_1 &=& 5\rho + (5m-6n)\,\frac{M}{2} \; .
\end{eqnarray}
Since $5m-3n$ and $5m-6n$ are arbitrary integers, we get
\begin{equation}
 5 A_1 ~=~ 5 \rho \mod \eta\;.  \label{eq:B2F}
\end{equation}

\subsection{Anomaly coefficients for $\boldsymbol{\Z{M}^R}$ symmetries}

Now consider a $\Z{M}^R$ symmetry, under which, by convention, the superpotential
transforms as
\begin{equation}
 \mathscr{W}~\to~\mathrm{e}^{2\pi\,\I\,q_\mathscr{W}/M}\,\mathscr{W}
\end{equation}
with $q_\mathscr{W}=2$. Accordingly, the superspace coordinates transform as
\begin{equation}
 \theta~\to~\mathrm{e}^{2\pi\,\I/M}\,\theta\;,
\end{equation}
such that $\D^2\theta$ transforms oppositely to $\mathscr{W}$. Superfields
$\Phi^{(f)}=\phi^{(f)}+\sqrt{2}\,\theta\psi^{(f)}+\theta\theta\,F^{(f)}$ follow
the law
\begin{equation}
 \Phi^{(f)}~\to~\mathrm{e}^{2\pi\,\I\,q^{(f)}/M}\,\Phi^{(f)}\;.
\end{equation}
Correspondingly, the fermions transform as
\begin{equation}
 \psi^{(f)}~=~\mathrm{e}^{2\pi\,\I\,(q^{(f)}-1)/M}\,\psi^{(f)}\;.
\end{equation}

For discrete $R$ symmetries, the anomaly coefficients read (\cf\
\Apref{app:PathIntegral})
\begin{subequations}
\begin{eqnarray}
 A_{G-G-\Z{M}^R}
 & = &
 \sum_{\boldsymbol{r}^{(f)}} 
 \ell\left(\boldsymbol{r}^{(f)}\right)\cdot(q^{(f)}-1)+\ell(\adj G)
 \;,\label{eq:A_G-G-ZNR}\\
 A_{\U1-\U1-\Z{M}^R}
 & = &
 \sum_m (Q^{(m)})^2\,\cdot(q^{(m)}-1)
 \;,\label{eq:A_U1-U1-ZNR}\\
 A_{\mathrm{grav}-\mathrm{grav}-\Z{M}^R}
 & = &
 -21+\sum_G \dim(\adj G)+\#(\U1)+\sum_m (q^{(m)}-1)
 \;.\nonumber\\
 & & \label{eq:A_grav-grav-ZNR}
\end{eqnarray}
\end{subequations}
Here $q^{(f)}$ denote the $\Z{M}^R$ charges of the superfields, the charges of
the corresponding fermions are shifted by one unit, $q_{\psi^{(f)}}=q^{(f)}-1$.
In \Eqref{eq:A_G-G-ZNR} $\ell(\adj G)=C_2(G)$ represents the contribution from
the gauginos, $\#(\U1)$ denotes the number of \U1 gauginos. The first and second
term on the right--hand side of \Eqref{eq:A_grav-grav-ZNR} represent the
contributions from the gravitino and gauginos. A necessary condition for anomaly
cancellation is the universality
\begin{eqnarray}
 A_{G-G-\Z{M}^R}
 & = &
 A_{\U1-\U1-\Z{M}^R}~=\frac{1}{24}~A_{\mathrm{grav}-\mathrm{grav}-\Z{M}^R}
 \nonumber\\
 & = & \rho\mod
 \eta\;.
 \label{eq:anomaly-universality-R}
\end{eqnarray}
$\rho$ is a constant, which is related to the discrete shift
\eqref{eq:DiscreteShiftDilaton} of the axion via
$\rho=\pi\,M\,\Delta_{\mathrm{GS}}$.

\subsection{Summary of anomaly coefficients}

The anomaly coefficients are given by
\begin{subequations}\label{eq:GeneralAnomalyCoefficients}
\begin{eqnarray}
 A_{G-G-\Z{M}^{(R)}} 
 & = &
 \sum_{\boldsymbol{r}^{(f)}} \ell(\boldsymbol{r}^{(f)})\,(q^{(f)}-R)
 	+\ell(\text{adj}\,G) \cdot R\;, \\
 A_{\U1-\U1-\Z{M}^{(R)}} 
 & = & \sum_m (Q^{(m)})^2\,(q^{(m)}-R)\;,\\
 A_{\mathrm{grav}-\mathrm{grav}-\Z{M}^{(R)}} & = & 
 R\cdot\left[-21+\sum_G \dim(\adj G) +\#(\U1)\right]
 \nonumber\\
 & & {}+\sum_m (q^{(m)}-R)
 \;, 
\end{eqnarray}	
\end{subequations}
where we distinguish between discrete non--$R$ ($R=0$) and $R$ ($R=1$)
symmetries. $\#(\U1)$ denotes the number of \U1 gauginos. As discussed above,
the  mixed $\U1-\U1-\Z{M}^{(R)}$ anomaly is only meaningful if one knows the
normalization. In general, the coefficient $A_{\U1-\U1-\Z{M}^{(R)}}$ is not
invariant under shifts of the $\Z{M}^{(R)}$ charges by integer multiples of $M$.

\section{A comment on K\"ahler stabilization}
\label{app:KaehlerStabilization}

A possible way to stabilize the dilaton is through non--perturbative 
corrections to the K\"ahler potential \cite{Shenker:1990uf,Banks:1994sg}. Such
corrections are expected to vanish in the limit of  zero coupling and also to
all orders in perturbation theory.  The form of these corrections has been
studied in the literature \cite{Casas:1996zi,Binetruy:1996nx,Binetruy:1997vr}. 
For a favourable choice of the  parameters, this correction may allow one to
stabilize the dilaton at a realistic value, $\re S\simeq 2$, while breaking
supersymmetry 
\cite{Casas:1996zi,Binetruy:1996nx,Binetruy:1997vr,Barreiro:1997rp,Buchmuller:2004xr}.
A common parametrization of the non--perturbative corrections reads
\begin{eqnarray}
 \mathrm{e}^K & = & \mathrm{e}^{K_0}+ \mathrm{e}^{K_\mathrm{np}} \;, \\
 \mathrm{e}^{K_\mathrm{np}} & = & c\, x^{p/2}\, \mathrm{e}^{-q\, \sqrt{x}}\;,
 \label{eq:Knp}
\end{eqnarray}
with $K_0=-\ln(2x)$, $x=\re S$, and parameters subject to $K'' > 0$ and 
$p,q > 0$. 
Supersymmetry is broken spontaneously by the $F$ term of the dilaton,
\begin{equation}
 F_S~\sim~\frac{\langle \lambda \lambda \rangle}{M_\mathrm{P}}\;.
\end{equation}
For a single gaugino condensate, one has
\begin{equation}
 \mathscr{W}~=~d\,\exp{\left(-\frac{3S}{2\beta}\right)}\;,
\end{equation}
where $3/(2\beta)=8\pi^2/N$ and $d=-N/(32 \pi^2\,\mathrm{e})$ for a condensing
$\mathrm{SU}(N)$ group with no matter. Note that the scalar potential
is independent of $\im S$. That is, we are left with a GS axion.

\begin{figure}[h!]
\centerline{
\includegraphics{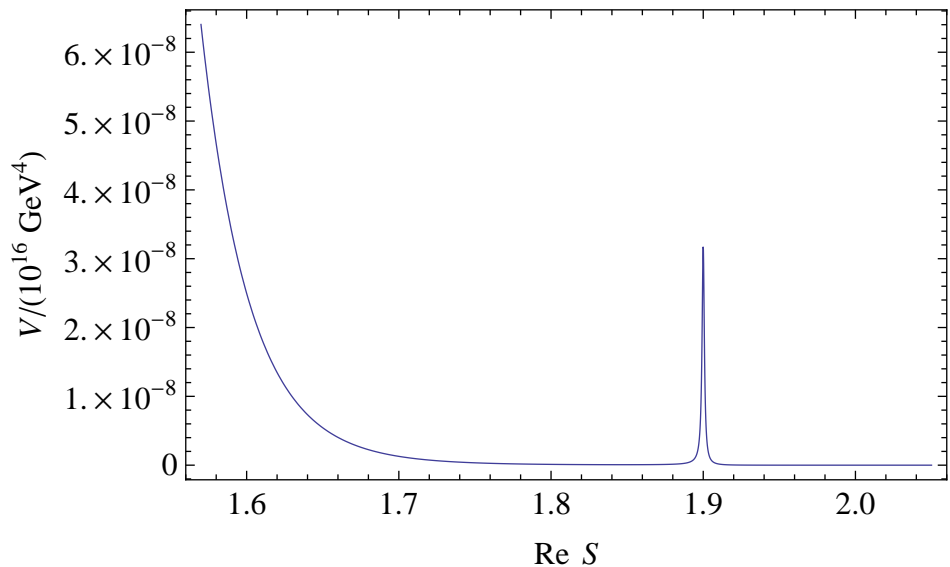}}
\caption{Dilaton potential for
$\{c_1,c_2,p_1,p_2,q\}=\{-28.8292,22.6129,2,3,4\}$.}
\label{fig:KaehlerStabTuned}
\end{figure}

The problem with this scheme is that the vacuum energy at the local minimum is
typically positive (\cf\ \cite{Barreiro:1997rp}). 
Although \Eqref{eq:Knp} represents the `standard' choice of the K\"ahler
potential, there are no arguments that forbid additional terms of the same
structure. That is, following Shenker's arguments \cite{Shenker:1990uf} one may
replace \Eqref{eq:Knp} by
\begin{equation}\label{eq:Knp2}
 \mathrm{e}^{K_\mathrm{np}} ~=~ 
 \left(c_1\, x^{p_1/2}+c_2\, x^{p_2/2}\right)\, \mathrm{e}^{-q \sqrt{x}}\;.
\end{equation}
In fact there seems to be no reason for not writing even more terms in the
parentheses. One can then tune the vacuum energy in the local minimum to zero by
carefully adjusting the coefficients (\Figref{fig:KaehlerStabTuned}).

Of course there is still no reason for why the vacuum energy should vanish at
the local minimum, but the above arguments may show that, in principle, the
vacuum energy can be tuned to zero in this scheme.

\bibliography{Orbifold}
\bibliographystyle{NewArXiv}

\end{document}